\documentclass{article}
\usepackage{graphicx} 
\usepackage{amsmath}
\usepackage{amssymb} 
\usepackage{hyperref}
\hypersetup{colorlinks}
\usepackage{fullpage} 
\usepackage{xcolor} 
\usepackage{pagecolor} 
\usepackage{tikz} 
\usepackage{optidef}
\usepackage{alphalph} 
\usepackage{comment} 
\usepackage{array} 
\usepackage{subcaption} 
\usepackage{soul} 
\usepackage{multirow}
\usepackage{multicol}
\usepackage{booktabs}
\usepackage{tabularx}

\newcommand\useStepCnt[1]{\the\value{#1}\stepcounter{#1}}
\newcommand\hidetxt[1]{}
\newcommand\deponly[1]{#1}
\newcommand\disableFig[1]{#1}
\renewcommand{\bold}[1]{\textbf{#1}} 
\newcommand{\fairness}[1]{}

\newcommand{\strevcnf}[1]{\st{}} 
\newcommand{\strev}[1]{\st{}} 
\newcommand{\streva}[1]{\st{}} 
\newcommand{\revcnf}[1]{\textcolor{black}{#1}}
\newcommand{\rev}[1]{\textcolor{black}{#1}}
\newcommand{\reva}[1]{\textcolor{black}{#1}}

\newcommand{\verti}{vertiminal}

\newcommand\formatone[1]{#1}
\newcommand\formattwo[1]{\hidetxt{#1}}

\newcommand{\anon}[1]{}
\newcommand{\nonanon}[1]{#1}

\title{Vertiport Terminal Scheduling with Multiple Surface Directions along with Throughput Analysis}
\author{Ravi Raj Saxena, T.V. Prabhakar, Joy Kuri, Manogna Yadav}
\date{May 2023}

\begin{document}
\maketitle

\begin{abstract}
Vertical Take-Off and Landing (VTOL) vehicles are gaining traction in both the delivery drone market and passenger transportation, driving the development of Urban Air Mobility (UAM) systems. 
UAM seeks to alleviate road congestion in dense urban areas by leveraging urban airspace. 
To handle UAM traffic, vertiport terminals (vertiminals) play a critical role in supporting VTOL vehicle operations such as take-offs, landings, taxiing, passenger boarding, refueling or charging, and maintenance. 
Efficient scheduling algorithms are essential to manage these operations and optimize vertiminal throughput while ensuring safety protocols.
Unlike fixed-wing aircraft, which rely on runways for take-off and climbing in fixed directions, VTOL vehicles can utilize multiple surface directions for climbing and approach. 
This flexibility necessitates specialized scheduling methods. 
We propose a Mixed Integer Linear Program (MILP) formulation to holistically optimize vertiminal operations, including taxiing, climbing (or approach) using multiple directions, and turnaround at gates. 
The proposed MILP reduces delays by up to 50\%.
Additionally, we derive equations to compute upper bounds of the throughput capacity of vertiminals, considering its core elements: the TLOF pad system, taxiway system, and gate system.
Our results demonstrate that the MILP achieves throughput levels consistent with the theoretical maximum derived from these equations. 
We also validate our framework through a case study using a well-established vertiminal topology from the literature.
Our MILP can be used to find the optimal configuration of vertiminal.
This dual approach, MILP and throughput analysis, allows for comprehensive capacity analysis without requiring simulations while enabling efficient scheduling through the MILP formulation.

\end{abstract}

\section{Introduction}
\label{intro}

Future modes of transportation in domestic mobility are expected to integrate Urban Air Mobility (UAM) with surface transport. 
This acceleration is due to road congestion caused by an increase in the number of vehicles for personalized transport. 
Frequent traffic rerouting can cause unpredictable delays either because of construction work, road accidents or due to ad-hoc events such as public and political demonstrations, etc. 
UAM is expected to provide cost-effective air travel by deploying fuel-efficient small passenger aircraft with electric mobility\rev{, popularly being called ``air taxi or air shuttle" service~\cite{9447255}}. 
The system is expected to be fully automated and, thus, highly efficient in operational procedures \cite{nasaConOps}.

\hidetxt{
Urban Air Mobility (UAM) is a convenient mode of future domestic transportation in large smart cities for all future domestic transportation.
UAM envisages a future in which advanced technologies and new operational procedures enable practical, cost-effective air travel as an integral mode of transportation in smart cities. 
It represents one of the most exciting and complex Advanced Air Mobility concepts, with highly automated aircraft providing commercial services to the public over densely populated smart cities.}

UAM vehicles, owing to their small size and primarily rotary wing structure, \streva{are equipped with the ability to}can take off and land vertically, making them ideally suited for widespread \streva{usage} \reva{operation} in congested and space-constrained \reva{urban} cities. 
The urban air traffic constituting such UAM vehicles, along with other Unmanned Aerial Systems (UAS) such as delivery drones, are governed by  UAS Traffic management or UTM. 
UTM mandates that UAM vehicles land and take off only at specified vertiports (vertical spaces) constructed inside city limits. 
\streva{UAM is projected as above \$40 billion industry.
Especially in Europe, companies such as Volocopter and Skyroads, in Italy and Germany, are already developing Vertical Take-Off and Landing (VTOL) technology, vertiports and air traffic automation technology.}
\reva{The UAM industry is projected to exceed \$40 billion in value~\cite{McKinsey}, with significant advancements already underway in Europe. 
Companies like Volocopter and Skyroads in Italy and Germany are actively developing Vertical Take-Off and Landing (VTOL) technologies, vertiports, and automated air traffic systems~\cite{volo1,volo2,skyroad}.}

\subsubsection*{Vertiport Terminal --- distinct from vertiport on rooftop}
\rev
{
A vertiport terminal consists of several Touchdown and Lift-Off (TLOF) pads and gates with taxiways connection, and it occupies a significantly large area footprint. Some of the vertiport terminals also consist of staging stands for overnight parking or maintenance of UAM vehicles compared to a typical rooftop vertiport. 
Whereas, a rooftop vertiport occupies a smaller area and would consist of one or two TLOF pads\hidetxt{ and few gates}~\cite{9441631}.
These vertiports function as individual stops, analogous to bus stops.
Vertiport terminals (or vertiminals), on the other hand, serve as central hubs for UAM operations, comparable to bus terminals in ground transportation systems.}
Designing vertiminals for a city is a nontrivial task as there is a requirement for efficient space usage as well as adhering to\hidetxt{along with maintaining} regulations. 
Several international aviation bodies such as EASA \cite{EASADoc}, FAA \cite{FAAdoc}, and UAE General Civil Aviation Authority\cite{UAE} have recently published design guidelines for the construction of vertiports and vertiminals.

\reva{
A notable analogy exists between Urban Air Mobility Traffic Management (UTM) and conventional Air Traffic Management (ATM).
\hidetxt{The capacity and demand of the UTM system in a city can be analogous to the high traffic volume observed between major airports of metropolitan areas~\cite{Watkins9473838, mueller2017enabling}.}
Vertiport terminals, akin to airports, are expected to become critical bottlenecks in UTM under high traffic conditions~\cite{Watkins9473838, mueller2017enabling}, leading to congestion and associated wastage of resources such as fuel (energy), manpower, and time, and thereby leading to uneconomical operation.
Hence, optimizing vertiminal operations is vital in intelligent transportation research.
This work draws upon well-established optimization methodologies from ATM research~\cite{jiang2015taxiing, lee2012comparison, deng2020novel} to enhance vertiminal terminal efficiency.
While existing studies on vertiports and vertiminals have primarily focused on terminal design ~\cite{vazquez2021vertiport, zelinski2020operational} and approach operations~\cite{shao2021terminal, song2021development}, limited attention has been given to ground operations.
Efficient ground operations including taxi route and TLOF pad selection, integrated taxiing and TLOF pad scheduling, and gate assignment (detailed in Section~\ref{probelm}) are crucial for minimizing resource wastage and improving throughput. 
We define the throughput of a system as the number of UAM vehicles arriving and/or departing in a time window. 
}

Unlike traditional aircraft, which face bottlenecks at runways~\cite{furini2015improved}, VTOL aircraft use TLOF pads for take-off and soon after are free to climb in any direction. 
This flexibility of climbing in any direction can be leveraged to design efficient schedules to increase the throughput while adhering to safety constraints such as wake vortex and minimum separation distance requirements.
However, these constraints introduce delays across various operations such as gate departure, taxiing, takeoff, and climbing.
To address these scheduling challenges, we propose a Mixed Integer Linear Program (MILP) formulation that optimally schedules vertiminal operations by minimising the sum of weighted delays.
The weight assignments offer operational flexibility, allowing operators to prioritise tasks as needed. A few examples from this formulation include: (a) prioritising arrivals over departures (b) prioritising congestion on taxiways over surface directions.  


\nonanon{Building on our previous work}\anon{Building on the work done by}~\cite{saxena2023integrated}, this paper extends the MILP framework by incorporating arriving flights, modelling gate turnaround operations, and significantly reducing computational complexity.
In addition, we derive analytical throughput bounds, offering deeper insights into vertiminal operations and capacity limitations.
We further validate our MILP formulation and throughput analysis using the Gimpo Vertiminal case study~\cite{ahn2022design}.
The main contributions of this paper are as follows:

\begin{enumerate}
    \item \hidetxt{We have proposed} A novel operational procedure is proposed where a UAM flight can use multiple climb and approach surface directions. \reva{Using the proposed MILP for scheduling, the delay is reduced up to 50\%, increasing the throughput of the vertiminal.}\streva{ This results in overall delay reduction (upto 50\%) and an increase in throughput.}
    
    \item The MILP model \hidetxt{presented in our previous work. We have further enhanced the MILP problem from our previous work for } is extended to incorporate arriving flights and turn around at gates. We reduce the number of variables, decreasing the computation time significantly by more than 90\% \strev{reduce the number of variables \& constraints, }. Comparison with First Come First Serve (FCFS) scheduling highlights the benefits of our approach.
    \strev{and have compared our formulation with First Come First Serve (FCFS) scheduling.} 
    
    \fairness{ \item To ensure a uniform delay distribution for UAM vehicles, we introduce a fairness metric.}
    
    \item We \streva{provide an approach}\reva{derive equations} to compute bounds on the throughput of a vertiminal that considers various sequences of aircraft movements and different flight classes. The calculations using these equations show that the upper bound on throughput obtained from the MILP scheduling is tight. 
    \item As a case study, we adopt one of the most optimized vertiminal topologies proposed in~\cite{ahn2022design}. 
\end{enumerate}
The rest of the paper is organized as follows. 
Section \ref{related} is on related work in literature. 
Formulation and explanation of the optimization problem \formattwo{along with its results} are presented in Section \ref{probelm}. 
Section \ref{sec: tc} derives the equations required to calculate the throughput of a \verti.
\formattwo{We have also compared the results of throughout calculation with the MILP formulation.}
\formatone{The computational results of the optimization problem and the throughput calculations, along with their comparison, are explained in Section~\ref{sec: results}.}
The case study of our work is presented in Section~\ref{sec: Case Study}.
Finally, in Section \ref{conclusion}, we summarise our work and discuss the future scope of the study.

\section{Related Work}
\label{related}
\subsection{\rev{Ground operations in conventional airports}}
Taxiing and runway scheduling is a well-researched area for traditional aircraft and airports. 
Numerous works like (\cite{gotteland2003genetic,jiang2015taxiing,liu2010airport}) have developed genetic algorithm-based techniques for solving taxiing problems. 
These works calculated a fit function based on taxiing length, delays, conflicts, etc. 
\streva{Genetic algorithms are computationally faster than an optimisation solver, but provide suboptimal solutions.} 
\reva{While computationally efficient, genetic algorithms often yield suboptimal solutions compared to exact optimisation methods such as MILP.}
The framework presented in~\cite{clare2011optimization} describes a  MILP method for joint \hidetxt{the coupled problems of} airport-taxiway routing and runway scheduling, \reva{using a receding-horizon approach to enhance scalability}. \streva{The receding-horizon formulation is applied for scalability.} 
\reva{Similarly, the work presented in~\cite{lee2012comparison} introduced MILP and a sequential approach to solve runway - taxiway scheduling.}
\streva{Work presented in cite{lee2012comparison} suggests two strategies for jointly optimising taxiway and runway schedules. While the first one uses an integrated MILP model, the second strategy sequentially integrates the runway and taxiway scheduling algorithms.}
The thesis work of Simaiakis \cite{simaiakis2013analysis} analyses the departure process at an airport using queuing models and proposes dynamic programming algorithms. 
For aircraft arrivals, a novel approach described in \cite{cheng2014airport} selects  the appropriate runway exit, which is the outcome of the taxiing scheduling problem that considers uncertain runway exit time and past exit selection patterns.  
\reva{In recent advancements, machine learning techniques have been employed to optimise ground operations.}
Furthermore, the work in \hidetxt{Further work such as}~\cite{herrema2019machine} has used Machine Learning (ML) techniques to determine runway exit, and its inference is utilised in~\cite{morgan2019validation} \hidetxt{has included these studies}for calculating runway utilisation and validation.
\reva{Gate assignment has also been explored as a critical component of airport scheduling as they affect runways to taxiway routes. Studies such as~\cite{behrends2016aircraft,deng2020novel,deng2017study}have examined gate assignment problems, focusing on passenger-centric metrics.}
\streva{Under the large context of airport scheduling, the works in cite{behrends2016aircraft,deng2020novel,deng2017study} consider the gate assignment problem as an essential part of the taxiing route from runway to the gate and vice-versa, but they are either passenger-centric or use a genetic algorithm.}
\strev{However, they are either passenger-centric or use a genetic algorithm.}
\reva{In summary, while the works in literature  address various aspects of airport ground operations, they typically treat arrival, turnaround, and departure processes in isolation, \textit{but none of the works optimise the operations holistically.}}
\hidetxt{\textit{none of these works holistically optimises the operations}}


\subsection{\rev{Vertiport and Vertiminal design and operations}}
Recent works \hidetxt{have concentrated }on vertiport scheduling and design such as \cite{kleinbekman2018evtol} present the methods to determine optimal ``Required time of arrival" using MILP by considering energy and flight dynamics, while \cite{shao2021terminal} proposes a multi-ring structure over a multi-vertiport. 
The multi-ring structure only handles air traffic \reva{by constructing control rules of junctions based on the cyclic phase backpressure strategy and an integrated adaptive scheduling model based on traffic structure and operation control rules. However, it} does not consider handling ground operations.
The thesis work in \cite{vazquez2021vertiport} discusses the software tools for vertiport design simulation and analysis.
Since a vertiminal can have multiple gates and TLOF pads, several works such as \cite{vascik2019development,ahn2022design,preis2022vertiport, zelinski2020operational} have conducted extensive studies on their design\streva{of vertiport and its}, capacity and throughput analysis.\hidetxt{They have put forth four topologies of vertiport designs: Linear, satellite, pier and remote apron}
The work presented in \cite{vascik2019development} reviewed various heliport topologies (linear, satellite, pier, and remote apron) for vertiminals and developed an integer program based on the Bertsimas-Stock multi-commodity~\cite{bertsimas2000traffic} flow model to analyse vertiminal operations. Their extensive experiments investigated the impact of factors like gate-to-TLOF pad ratio, operational policies, and staging stand availability on throughput and capacity envelope. However, they have not considered the inter-separation distance requirements on taxiways and surface directions\hidetxt{or wake vortex separation requirement on the TLOF pad} among different classes or types of flights.
The study in \cite{ahn2022design} analysed various topologies for maximising throughput at Gimpo airport vertiminal. These topologies are based on EASA and FAA guidelines. 
While the authors have used obstacle-free volume, they do not explore the concept of multiple surface directions.
The work of \cite{preis2022vertiport} proposed a heuristic for vertiminal design focusing on efficient ground operations in UAM. Their work analysed average passenger delay based on vertiminal layout and demand profile throughout the day.
\rev{\textit{In summary, none of these works has considered multiple climb and approach surface directions for vertiminal operations.}}

\subsection{\rev{Our Contributions}}
To the best of our knowledge, no work has considered multiple climb and approach directions for take-off and landing, \hidetxt{along}with an integrated approach towards ground operation processes such as taxiing and turnaround at gates at a \streva{vertiport terminal, termed as} vertiminal.
\streva{Also, no work has studied and analysed the impact on the throughput of vertiminal by arrival and departure operations, considering different sequences of operations such as arrival-arrival, arrival-departure, etc., with different classes or types of VTOL vehicles.}
\reva{The impact of arrival and departure operations on vertiminal throughput, particularly across different operational sequences (e.g., arrival-arrival, arrival-departure) and various classes or types of UAM vehicles, remains largely unexplored. Our work offers fresh insights into these aspects.}
Our work builds upon the optimisation problem proposed in \cite{tsao2009integrated}. Unlike the base formulation which minimises the weighted sum of the gate, taxiing, and queuing delays\hidetxt{ in operations} at conventional airports, our work minimises the weighted sum of delays for approach, landing, taxiing, turnaround, gate, take-off, and climbing operations at vertiminals.
\hidetxt{
The work models runway entrance as a queue to \hidetxt{account for the inability to }adhere to the schedule developed by the deterministic optimisation process.
Unlike their work, we are not using queues at the entrance of the TLOF pad. }
\hidetxt{Having different directions to climb, the take-off can be free-flowing, and no aircraft needs to wait at the entrance of the TLOF pad, assuming proper aircraft scheduling.
The impact of queuing, if any, is left for future study. }

\section{Vertiminal Operations}
\label{probelm}

\begin{figure}[htb] 
  \centering
  \begin{subfigure}[b]{\columnwidth}
    \centering
    \disableFig
    {\includegraphics[scale=0.25]{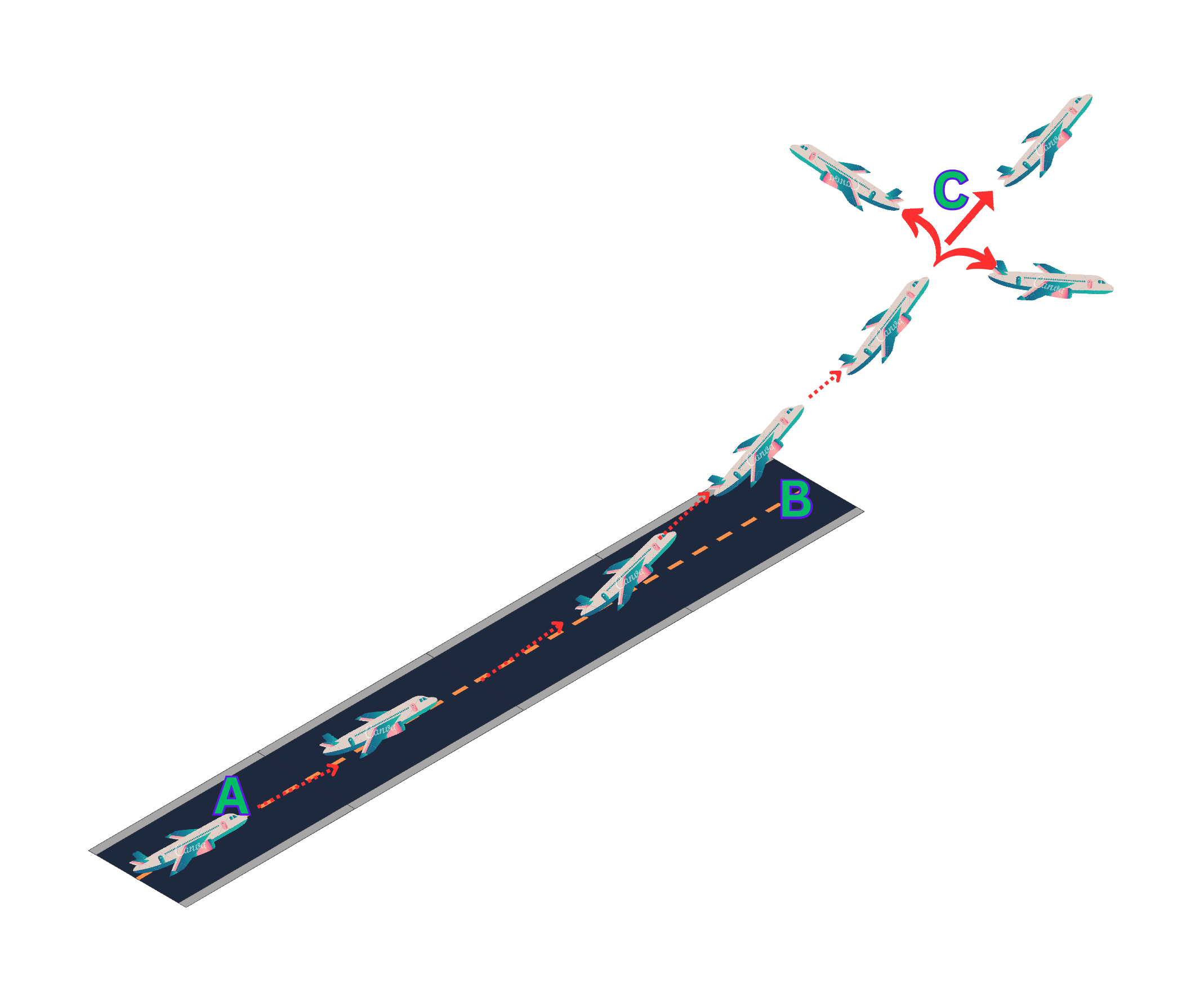}}
    \caption{Aircraft flight path}
    \label{fig:air path}
  \end{subfigure}
  \begin{subfigure}[b]{\columnwidth}
    \centering
    \disableFig
    {\includegraphics[scale=0.06]{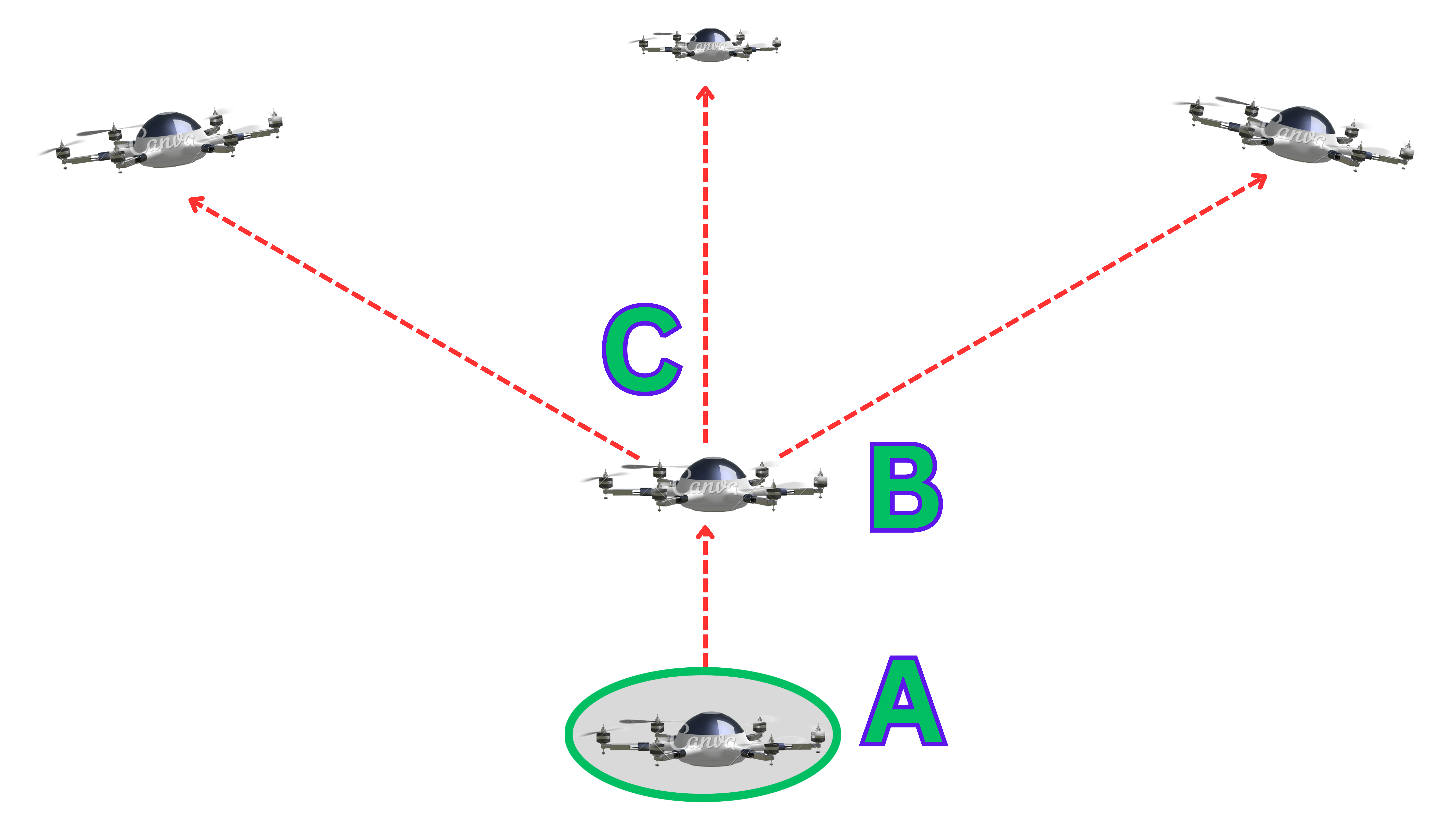}}
    \caption{VTOL vehicle flight path}
    \label{fig:drone path}
  \end{subfigure}
  
  \caption{Comparison of flight paths}
  \label{fig:flight_paths}
\end{figure}
The ground operations on a vertiminal are similar to those of an airport, however, the difference lies in their arrival or departure operations.
Figure \ref{fig:air path} shows a conventional airport system where an aircraft occupies the runway, accelerates to the required ground speed, take-off, and after attaining a certain altitude will deviate in the direction of the \strevcnf{destination} \revcnf{approved flight plan}\hidetxt{next phase of its flight}. The two line segments ``A to B" and ``B to C" depict the \strevcnf{minimum common distance travelled by any aircraft} \revcnf{common distance occupied by only one flight at a time, thus being a bottleneck to back-to-back \hidetxt{the following}departing traffic.}

\strev{We define \textbf{vertiminal} as a vertiport terminal, which may be regarded as similar to a bus terminal, and a vertiport is like a bus stop.}
A \rev{\bold{vertiminal}} handles traffic of \textbf{UAM VTOL capable aircraft}. 
Such aircraft are defined as heavier-than-air aircraft, other than an aeroplane or helicopter, capable of performing VTOL using more than two lift or thrust units to provide lift during the take-off and landing \cite{EASADoc}.
\revcnf{A UAM aircraft departs from the \textbf{Gate} and travels a defined path on vertiminal called a \textbf{taxiway}, either on the ground or by air, before reaching the \textbf{TLOF pad}.}
Unlike conventional runways, these pads are compact, and thus, a VTOL vehicle can reach cruising altitude (point C) quickly \hidetxt{and starts its next phase of flight} \revcnf{and start its approved flight plan}.
Figure \ref{fig:drone path} shows a VTOL vehicle take-off from a {TLOF pad}. The line segment ``A to B" shows the vertical take-off
and \hidetxt{unlike in Figure~\ref{fig:air path},}the line segments ``B to C" is of $0$ length.

\begin{figure}[ht]
    \centering
    \disableFig
    {
    \includegraphics[width=\linewidth]{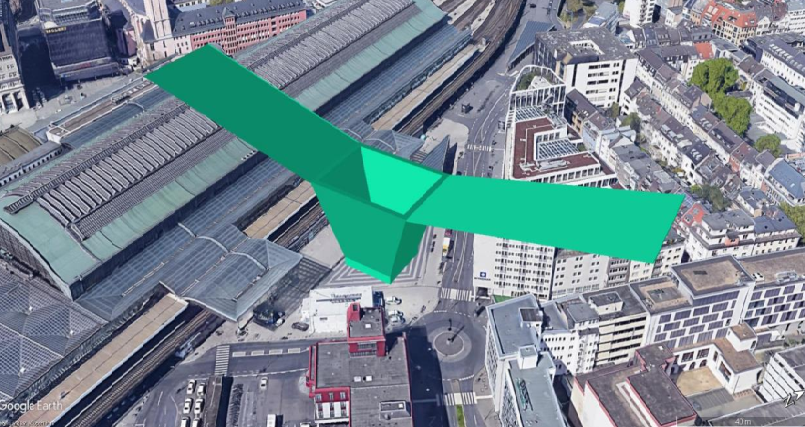}
    }
    \caption{Example of OFV with 2 \deponly{approach and } climb surface directions \strevcnf{(Courtesy Figure D22 of )} \cite{EASADoc}.}
    \label{fig:OFV}
\end{figure}

Figure~\ref{fig:OFV} shows an example of \textbf{Obstacle Free Volume} (OFV), which is a funnel-shaped area with several climb surface directions (2 in Fig~\ref{fig:OFV}) available to\hidetxt{present on} the TLOF pad. The OFV
guarantees that VTOLs can accomplish take-offs and landings within a sizable vertical segment, designed suitably to remain within \hidetxt{allowing them to account for}environmental and noise limits in urban settings. 
At the end of the OFV boundary, there can be multiple climb surface directions for VTOL vehicles to reach the vertiexit.
We define \textbf{vertiexit} as a point where the UAM VTOL aircraft exits from the vertiminal airspace at the cruising altitude.
\rev{For safe operations, only one UAM vehicle should occupy OFV.}
A minimum separation time is required for successive take-offs and landings to overcome wake turbulence on the TLOF pad.
\strevcnf{From Figure, point C shows the core idea where one may exploit several degrees of freedom available for a direction change and thus significantly reduce congestion and delays incurred by departing UAM aircraft.}
Two UAM aircraft \strevcnf{moving in}\hidetxt{flying in}climbing or approaching on the same surface direction \hidetxt{must be separated by the}are required to maintain a minimum separation distance. 
However, no minimum separation distance is required\hidetxt{ to be maintained} if the two aircraft \strevcnf{move in} fly in different surface directions. 
Hence,\hidetxt{ with more than one surface direction,} the total delay can be reduced\hidetxt{compared to a single surface direction}, provided the scheduling algorithm takes advantage of the multiple directions when deciding the take-off/landing sequence. 
The addition of multiple directions thus effectively reduces the shared path length and increases throughput.

{
A departure-ready flight can experience delays along its path due to the following:
\begin{enumerate}
    \item Separation requirements on the taxiways;
    \item TLOF pad availability while the previous VTOL aircraft climbs the OFV boundary;
    \item Wake turbulence time separation requirements on TLOF pad;
    \item Separation requirements on the climb surface direction in case multiple consecutive UAM vehicles take the same direction.
\end{enumerate}
Arriving flights will experience similar delays, however, in the reverse order. 
Our goal is to formulate a scheduling problem to minimise a weighted sum of all such delays that should help improve
the throughput of a vertiminal.
}

\subsection{Problem Formulation}
\label{sec: MILP}
\let\hidetxt\undefined
\let\deponly\undefined
\newcommand{\descp}[1]{\textcolor{purple}{}}
\newcommand{\hidetxt}[1]{}
\newcommand{\deponly}[1]{#1}

Given the number of UAM aircraft, their classes, their operation routes and their arrival \& desired departure times, we need to design an efficient scheduling algorithm. 
We use the MILP technique to formulate the scheduling problem.
The objective is to minimize the weighted sum of delays a UAM aircraft experiences during its complete operation.
The time consumed on \hidetxt{highly }constrained resources like TLOF pads would be penalized more than the less constrained resources such as gates.
The constraints for the objective problem are the physical constraint, separation requirements and other ATM regulations.
In Section~\ref{sec: MILP results}, we will discuss the effect of multiple surface directions on departure delay experienced by different flight loads\hidetxt{ numbers of flights } and their classes \hidetxt{in their operation} using the formulation presented in this section.
We will also compare our formulation with First Come First Serve (FCFS) scheduling.
Throughout this work, we use the term VTOL and flight interchangeably instead of UAM aircraft. 
We use the term ``surface direction'' in the optimization problem to denote both the take-off climb surface direction and approach surface direction.

\subsubsection{Assumptions}

As explained below, several simplifying assumptions are made in modelling vertiminal operations.
\begin{itemize}
    \item Vertiminals will provide a standard fixed-width surface direction to ensure VTOL remains within this specified limit.

    \item The approach or climbing speed of a VTOL  is assumed to be given and governed by equipment type, environment and regulations.
    

    \item \hidetxt{Vertiminals have standard taxi routes. Therefore,} Given a TLOF pad and a gate, the taxi route of each VTOL  is predefined.

    \item Nominal taxi speed is assumed to be given. Thus, given the length of the taxiway, the minimum and maximum travel time can be estimated. 



    \item The passage times at critical points along taxi routes, OFV and surface direction given by the solution to the optimization problem can be met by a VTOL. We also assume hovering is not allowed for VTOLs.
    \item For the flights that do not have to turn around, gates act as a source of flights and vertiexit act as a sink of flights for the departures. While for arrivals, vertiexit is a source of flights, and gates act as a sink.
    \item The number of VTOLs that have to turn around on the gates will not exceed the holding capacity of the gate.
    \item At the gates, for turnaround flights, the flight whose boarding completes first will be ready to leave first.

\end{itemize}

\subsubsection{VTOL  Operations in a \verti}
\begin{figure}[htbp]
        \disableFig
        {\includegraphics[width=0.8\linewidth]{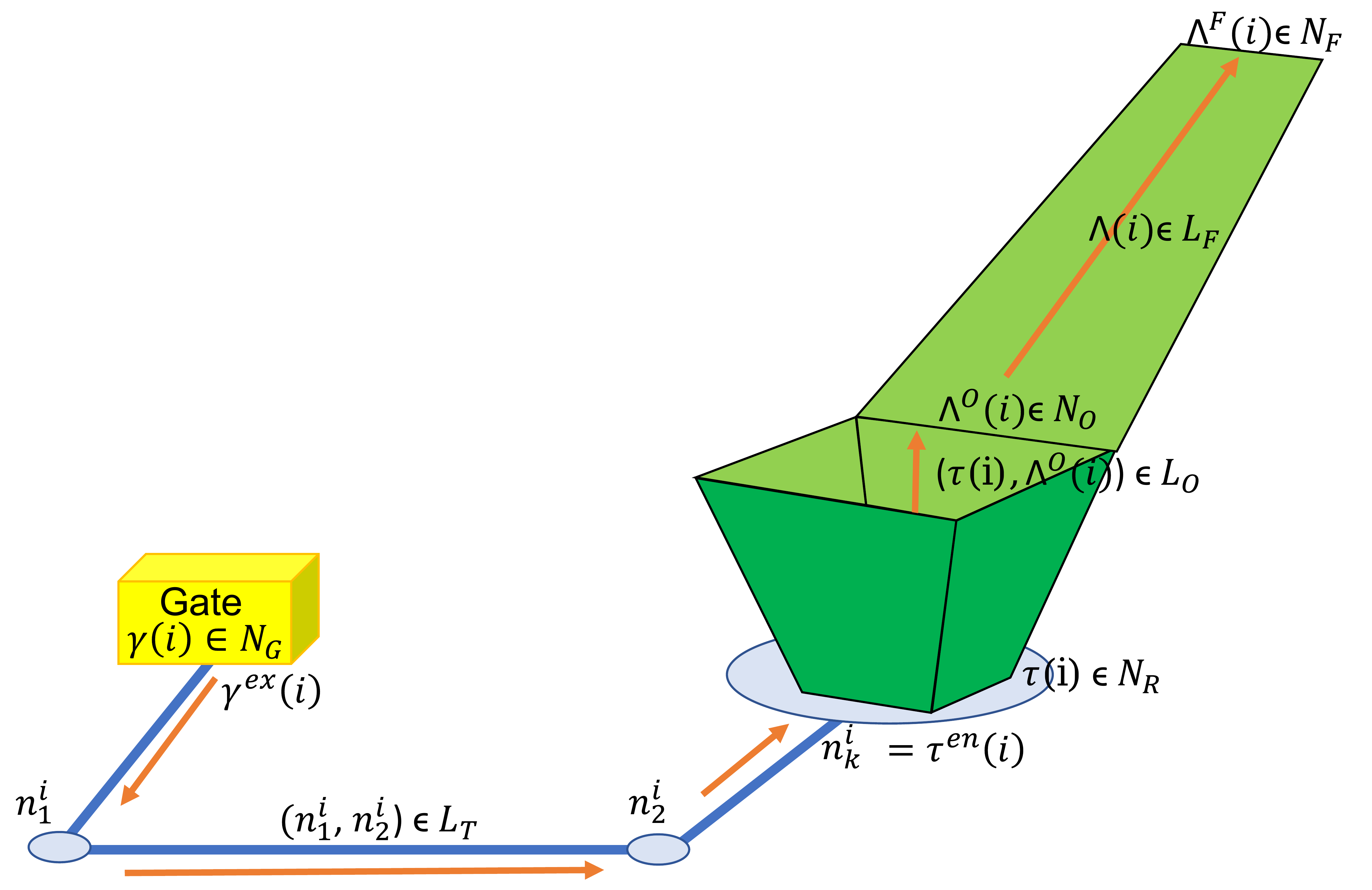}}
        \caption{The nodes and link representation for the optimisation problem that are explained in Table~\ref{tab: vert symbols}. The orange line represents the path of a VTOL from gate to vertiexit.}
        \label{fig:topo_Opt}
\end{figure}

     The Table \ref{tab: vert symbols} and Table \ref{tab: vtol sets} represent notation for a \verti\ and VTOL on a \verti, respectively. 
     Figure~\ref{fig:topo_Opt} shows the nodes and links explained in Table~\ref{tab: vert symbols}.
     Let $P_i\ \forall i \in A$ represent the physical taxi route for a VTOL  $i$, from its first node to its last node. 
    $P_i$ is represented by an ordered set of $k_i$ nodes $\{ n^i_1, n^i_2, ..., n^i_{k_i}\} \in N_T$ starting with the first node $n^i_1$ of the route, which is the exit of assigned TLOF pad for $i \in A^{Arr}$ and gate exit for $i \in A^{Dep}$, progressing through the intermediate nodes and ending at the last node of route $n^i_{k_i}$ which is the gate entrance for $i \in A^{Arr}$ or assigned TLOF pad entrance for $i \in A^{Dep}$, where $k_i$ denotes the total number of nodes involved in $P_i$. $P^a_i$ represents taxiing route $\forall i \in A^{Arr}$ while $P^d_i$ represent taxiing route $\forall i \in A^{dep}$.%

A VTOL  $i$ with a stop on the vertiminal will follow the following route:
\begin{enumerate}
\item Link $\Lambda^{Arr}(i)$  represents approach direction to the assigned TLOF pad $\tau^{Arr}(i)$.
\item Link connecting OFV boundary node $\Lambda^{Arr}_O(i)$ and TLOF  pad $\tau^{Arr}(i)$.
\item A sequentially ordered set $P^{a}_i$ of  nodes $n \in N_T$, from the TLOF pad exit node $\tau^{ex}(i)$ \hidetxt{on the network G }to the entrance of the gate $\gamma^{en}(i)$.
\item For the \textit{turnaround} flights\hidetxt{ which have to ,} $i \in A^{TAT}$, VTOL  would be in the holding space of the gate $\gamma(i)$\hidetxt{, from the entrance of gate $\gamma^{en}(i)$ to its exit $\gamma^{ex}(i)$}, where boarding and deboarding occur. 
\item A sequentially ordered set $P^{d}_i$ of  nodes $n \in N_T$, from the starting gate node $\gamma^{ex}(i)$\hidetxt{ on the network G} to the entrance of the assigned TLOF pad $\tau^{en}(i)$
\item Link connecting OFV boundary $\Lambda^{Dep}_O(i)$ and TLOF  pad $\tau^{Dep}(i)$.
\item Link $\Lambda^{Dep}(i)$  represents climb direction to the assigned TLOF pad $\tau^{Dep}(i)$.
\end{enumerate}


\begin{table}[htbp]
    \caption{Decision Variables}    \label{tab:decision_variables}
    \begin{tabular}{|l|m{0.95\textwidth}|}
        \hline
        $t^{n}_{i}$ & $ \forall i \in A, \forall n \in N$, Non-negative variable representing the time at which VTOL $i$ reaches node $n$. Note that the VTOL does not slow down, let alone stop, before reaching a node and continues to achieve smooth travel unless the node is the exit or entrance of a TLOF pad or gate. \\
        \hline
        $y^n_{ij}$ & $ \forall n \in N, \forall i,j \in A^n i \neq j$ Binary variable, indicating if VTOL $i$ reaches node $n$ before VTOL $j$ does. $y^n_{ij} = 1$ if and only if VTOL $i$ reaches node $n$ before VTOL $j$ does; $y^n_{ij} = 0$ otherwise. \\
        \hline
        $z_{ij}$ & $ \forall i,j \in A^{TAT}, i \neq j$ Binary variable used for selecting the minimum exit time for VTOLs in the Turnaround Time (TAT) set. \\
        \hline
    \end{tabular}
\end{table}

\newcommand{\tni}{t^n_i}
\newcommand{\tnj}{t^n_j}
\newcommand{\tri}{t^{\tau(i)}_i}
\newcommand{\trj}{t^{\tau(j)}_j}
\newcommand{\tria}{t^{\tau^{Arr}(i)}_i}
\newcommand{\trid}{t^{\tau^{Dep}(i)}_i}
\newcommand{\triex}{t^{\tau^{ex}(i)}_i}
\newcommand{\trien}{t^{\tau^{en}(i)}_i}

\newcommand{\gati}{t^{\gamma}_i(i)}
\newcommand{\gatj}{t^{\gamma}_i(j)}
\newcommand{\gaten}{t^{\gamma^{en}(i)}_i}
\newcommand{\gatex}{t^{\gamma^{ex}(i)}_i}

\newcommand{\yij}{y^n_{ij}}
\newcommand{\yji}{y^n_{ji}}

\newcounter{Constraint_count}
\stepcounter{Constraint_count}

\subsubsection{Formulation}
\reva{The objective function~\eqref{obj eqn} minimizes the weighted sum of delays incurred during departures, arrivals, and turnaround operations for VTOL vehicles. 
These delays are computed for each operational stage, with the decision variables and weights detailed in Tables~\ref{tab:decision_variables} and~\ref{tab:weight_symbols}, respectively. Below, the delay components for each operation type are outlined:}\\

\reva{
For departures ($i \in A^{Dep}$)}
\setlength{\leftmargini}{1.5cm}
\begin{itemize}
    \item waiting at the departure gate: $GD^D_i = {W_g}(\gatex - \mathtt{DRGATE}_i)$ 
    \item taxiing from gate exit to TLOF pad entry: $TD^D_i = W^{Dep}_t(\trien - \gatex)$ 
    \item climbing inside OFV: $OD^D_i = {W^{Dep}_r}(t^{\Lambda^{Dep}_O(i)}_i - \trien)$ 
    \item climbing to vertiexit from OFV boundary: $SD^D_i = {W_c^{Dep}}(t^{\Lambda^{Dep}_F(i)}_{i} - t^{\Lambda^{Dep}_O(i)}_i)$
\end{itemize}

\reva{
For arrivals ($i \in A^{Arr}$)}
\begin{itemize}
    \item approach from vertiexit to OFV boundary: $SD^A_i = {W_c^{Arr}}(t^{\Lambda^{Arr}_O(i)}_i - \mathtt{ARAPPR}_i)$ 
    \item leaving the TLOF pad: $OD^A_i = W^{Arr}_r(\triex - t^{\Lambda^{Arr}_O(i)}_i)$
    \item reaching the gate: $TD^A_i = W^{Arr}_t(\gaten - \triex)$
\end{itemize}

\reva{
For turnaround VTOLs ($i \in A^{TAT}$)}
\begin{itemize}
    \item turnaround time on the gates: $TTD_i = W_Q(\gatex - \gaten)$
\end{itemize}

\descp{check the definition of different weights to know about each term\\}

\begin{equation}
    \sum_{i \in A^{Dep}}[GD^D_i + TD^D_i + OD^D_i + SD^D_i] + \sum_{i \in A^{TAT}}TTD_i + \sum_{i \in A^{Arr}}[SD^A_i + OD^A_i + TD^A_i] \label{obj  eqn}
\end{equation}

\reva{The constraints are as follows:}\\
C\useStepCnt{Constraint_count}:
We impose constraint~\eqref{Arr C1} to ensure that an arriving VTOL  starts on the approach surface direction starting from vertiexit.
\begin{equation}
{t^{\Lambda^{Arr}_F(i)}_i}{= \mathtt{ARAPPR}_i \label{Arr C1}}{\qquad}{\forall i \in A^{Arr}}
\end{equation}

C\useStepCnt{Constraint_count}: A VTOL  $i \in A^{Arr}$ traverses physical nodes $p \in  \{\Lambda^{Arr}_F(i) \cup \Lambda^{Arr}_O(i) \cup \tau^{Arr}(i) \cup \ P^{a}_i \cup \gamma(i)\}$  while a VTOL  $i \in A^{Dep}$ traverses physical nodes $p \in  \{P^{d}_i \cup \tau^{Dep}(i) \cup \Lambda^{Dep}_O(i) \cup \Lambda^{Dep}_F(i)\}$.
For speed control and smooth travel over all the links,i.e. taxiing ways, OFV and surface direction, constraint~\eqref{Taxi and climb C6} is required.
The terms $T^{min}_{il^i_p}\ and\ T^{max}_{il^i_p}$ are the respective minimum and maximum time taken by VTOL  $i$ on the link $l^i_{p,p+1} \in L$.

\begin{equation}
\label{Taxi and climb C6}
     t^{p}_{i} + T^{min}_{il^i_{p,p+1}} \leq t^{p+1}_{i} \leq t^{p}_{i} + T^{max}_{il^i_{p,p+1}}
\end{equation}

C\useStepCnt{Constraint_count}: A VTOL  will spend time on the gate not less than its turn around time, which is ensured by the constraint~\eqref{TAT C7} 
\begin{equation}
    {\gatex - \gaten}{\geq TAT_i \label{TAT C7}}{\qquad}{\forall i \in A^{TAT}}
\end{equation}

C\useStepCnt{Constraint_count}: 
The constraint~\eqref{Dep gate C2} is the minimum time when the VTOL  should leave the gate and start taxiing. 

\begin{equation}
    {\gatex}{\geq \mathtt{DRGATE}_i \label{Dep gate C2}}{\qquad}{\forall i \in A^{Dep}}
\end{equation}

C\useStepCnt{Constraint_count}: Definition of Predecessors $y^n_{ij}$:
\begin{subequations}

\begin{equation}
{t^{n}_{j}}{\geq t^{n}_{i} - (1-y^n_{ij})M \label{y C9} }{\qquad}{\forall n \in N, \forall i,j \in A^n, i \neq j}
\end{equation}
\descp{The actual constraint is $(t^n_j - t^n_i)*y^n_{ij} > 0$ for $\yij$=1, no problem but for $\yij$= 0, $\tni \leq M$\\
if i=6, j=10,  $t^n_{10} > t^n_6 -200 => t^n_6 = 40, t^n_{10} = 10$ is valid. THIS IS AN ANOMALY. \textcolor{green}{Anomaly wont occur considering all the equations}}

\begin{equation}
    {\yij + \yji}{ = 1 \label{y C12.1}}{\qquad}{\forall n \in N, \forall i,j \in A^n, i \neq j}
\end{equation}
\descp{A more fundamentally strong constraint making only one of the order correct}
\end{subequations}
\\




C\useStepCnt{Constraint_count}: The following constraint prevents overtaking in terms of $y^n_{ij}$:

\begin{equation}
\begin{split}
    {y^n_{ij} - y^m_{ij}}{=0 \label{overtake C18} }{\qquad}\\
    {\forall i,j \in A,  i \neq j, \forall l_{n,m} \in (P_i\cup \Lambda(i)) \cap (P_j \cup \Lambda(j))}
    \end{split}
\end{equation}
C\useStepCnt{Constraint_count}: For preventing head-on collision of two VTOLs on a link $l_{n,m}$, following constraint is used:

\begin{equation}
\begin{split}
    {y^n_{ij} - y^m_{ij}}{=0 \label{collision C19} }{\qquad}\forall i,j \in A,  i \neq j, \\
    \forall l_{n,m} \in (P_i \cup \Lambda(i) \cup L_O)\ and \ l_{m,n} \in (P_j \cup \Lambda(j) \cup L_O)
\end{split}
\end{equation}
C\useStepCnt{Constraint_count}: This constraint maintains the separation requirement on the Taxi and surface direction, where $L^{sep}_{ij}$ is the separation requirement between VTOL  in units of distance over the edge(or link) of length $L(n,m)$.

\begin{equation}
\begin{split}
    {t^{n}_{j}}{\geq t^{n}_{i} + \frac{L^{sep}_{ij}}{L(n,m)} (t^{m}_{i} - t^{n}_{i}) - (1- y^n_{ij})M \label{TaxiSep and climbSep C20} }\\
    {\forall l_{n,m} \in L_G \cup L_T\cup L_F \ \forall i,j \in A^n, i\neq j}
\end{split}
\end{equation}
\descp{Taking the ratio as $\rho$ , the actual constraint is $((\tnj - \tni) - \rho(t^m_i -t^n_i))\yij \geq 0$\\
No problem if $\yij=1$\\} 
C\useStepCnt{Constraint_count}: The constraint \eqref{Gate capacity} enforces the gate holding capacity of a gate $g \in N^G$ that should not be exceeded. We have shown for holding capacity of 3. 
~\eqref{Gate capcity1} linearize the constraint.
\begin{equation}
\begin{split}
    t^{g^{en}}_i >= min(t^{g^{ex}}_j, t^{g^{ex}}_k, t^{g^{ex}}_l) \\
    if \ y^g_{ji}=y^g_{ki}=y^g_{li}=1 \\
    g=\gamma(i)=\gamma(j)=\gamma(k)=\gamma(l),\\
    \forall i,j,k,l \in A^{TAT}, i \neq j \neq k \neq l
\end{split}
    \label{Gate capacity}
\end{equation}
\begin{subequations}
    \begin{equation}
        t^{g^{en}}_i \geq t^{g^{ex}}_j - (1-y^g_{ji})M - (1-z_{ij})M
    \end{equation}
    \begin{equation}
        t^{g^{en}}_i \geq t^{g^{ex}}_k - (1-y^g_{ki})M - (1-z_{ik})M
    \end{equation}
    \begin{equation}
        t^{g^{en}}_i \geq t^{g^{ex}}_l - (1-y^g_{li})M - (1-z_{il})M
    \end{equation}
    \begin{equation}
        z_{ij}+z_{ik}+z_{il} \geq 1 \label{Gate capacity1.4}
    \end{equation}
    \label{Gate capcity1}
\end{subequations}
C\useStepCnt{Constraint_count}: Constraint~\eqref{wake C22} maintains the wake vortex requirement on the TLOF pad when a VTOL  $i$ lands (or does a take-off) where $W^{tsep}_{ij}$ is the minimum time VTOL $j$ should wait to either arrive or depart after the arrival(or departure) of VTOL  $i$.
\begin{equation}
\begin{split}
    {t^{r}_{j}}{\geq t^{r}_{i} + W^{tsep}_{ij} - (1-y^{r}_{ij})M \label{wake C22} }\\
    {\forall i,j \in A^r, i \neq j, \tau(i)=\tau(j)=r}
\end{split}
\end{equation}
\descp{Original constraint $(\trj-\tri-W^{tsep}_{ij})y^r_{ij} \geq 0$. However similar anomaly exist as described in \eqref{y C9} \\}
C\useStepCnt{Constraint_count}: The time taken to reach the TLOF pad exit point upon landing is captured by constraint~\eqref{TLOF exit 21}.

\begin{equation}
    {\triex}{\geq  \tria + TOT^{Arr}_{i}\label{TLOF exit 21} }{\qquad}{\forall i \in A^{Arr}}
\end{equation}
C\useStepCnt{Constraint_count}: The time taken to take-off after entering the TLOF pad from an entry point is captured by constraint~\eqref{TLOF entrance 21}. Constraints~\eqref{TLOF entrance 21} and~\eqref{TLOF exit 21} along with~\eqref{Taxi and climb C6} maintain the time continuity over the VTOL's route. 

\begin{equation}
    {\trid}{\geq \trien + TOT^{Dep}_i\label{TLOF entrance 21} }{\qquad}{\forall i \in A^{Dep}} 
\end{equation}
\descp{}
C\useStepCnt{Constraint_count}: \hidetxt{To ensure that after the }Upon arrival of VTOL  $i$, the following VTOL  $j$ should reach the OFV boundary only after VTOL  $i$ has left the TLOF pad \hidetxt{and entered TLOF pad entrance or}through its exit point. Constraint~\eqref{land approach Arr C26.2} satisfies this requirement 
\begin{equation}
\begin{split}
    {t^{\Lambda^{Arr}_O(j)}_{j}}{\geq \triex - (1-y^r_{ij})M \label{land approach Arr C26.2} }\\
    {\forall i,j \in A^{Arr} , i \neq j}{,}{}{\tau^{Arr}(i) = \tau^{Arr}(j) = r}
\end{split}
\end{equation}
\descp{Original constraint $(t^{\Lambda^O(i)}_{j} -  t^{n^i_1}_{i} )y^r_{ij} \geq 0$. However similar anomaly exist as described in \eqref{y C9} \\}
C\useStepCnt{Constraint_count}: \hidetxt{Similar is the case }When a VTOL $i$ departs\hidetxt{; after the departure of a VTOL  } from the assigned TLOF pad, its immediate follower VTOL $j$ cannot enter the TLOF pad till VTOL $i$ \hidetxt{the departing VTOL  }has crossed its OFV boundary. Constraint~\eqref{takeOff climb C26.1} ensures this.
\begin{equation}
\begin{split}
    {t^{\tau^{en}(j)}_{j}}{\geq t^{\Lambda^{Dep}_O(i)}_i  - (1-y^r_{ij})M \label{takeOff climb C26.1} }\\
    {\forall i,j \in A^{Dep} , i \neq j}{}{}{, \tau^{Dep}(i) = \tau^{Dep}(j) = r}
\end{split}
\end{equation}
\descp{Original constraint $(t^{r}_{j} - t^{\Lambda^O(i)}_i)y^r_{ij} \geq 0$. However similar anomaly exist as described in \eqref{y C9} \\}
C\useStepCnt{Constraint_count}: A departing VTOL  must enter a TLOF pad only after the arriving VTOL  has left the TLOF pad. 
\begin{equation}
\begin{split}
    {t^{\tau^{en}(j)}_j }{\geq \triex - (1-y^r_{ij})M \label{noCollison C28}}\\
    {\forall i \in A^{Arr},j \in A^{Dep}, \tau^{Dep}(j) = \tau^{Arr}(i) = r}
\end{split}
\end{equation}

C\useStepCnt{Constraint_count}: Binary and non-negativity constraints:
\begin{subequations}
    \begin{equation}
        {t^{n}_{i}}{\in R^+ \label{var C30} }{\qquad}{\forall i \in A, \forall n \in N}
    \end{equation}
    \begin{equation}
        {y^n_{ij}}{\in \{ 0,1 \} \label{binvar C27} }{\qquad}{\forall n \in N, \forall i,j \in A^n_0, i \neq j}
    \end{equation}
\end{subequations}


\subsection{Comparison to the previous formulation}
\nonanon{In our earlier work~\cite{saxena2023integrated},we formulated the optimisation problem (\textit{Opt A}) using two binary variables: immediate predecessor $x$ and predecessor $y$.
Our present optimisation formulation (\textit{Opt B}) has eliminated the immediate predecessor variable $x$ because $y$ implicitly incorporates the information conveyed by $x$.} 
\anon{In the work done by ~\cite{saxena2023integrated}, the optimisation problem (\textit{Opt A}) uses two binary variables: immediate predecessor $x$ and predecessor $y$.
Our optimisation formulation (\textit{Opt B}) has eliminated the immediate predecessor variable $x$ because $y$ implicitly incorporates the information conveyed by $x$.}
This change eliminated the need for a dummy VTOL and also \streva{resulted in a reduction of}\reva{reduced} 10 constraint equations \hidetxt{associated with defining $x$ and $y$}and, thus, drastically reducing the time to reach the solution. 
\rev{Given the input parameters, the program generates all the equations for the solver to solve. 
Formulation time is the time the program takes to generate the equations, and solver time is the time the solver takes to solve the equations.
Table~\ref {tab: optAvsoptB} compares \textit{Opt A} and \textit{Opt B} in terms of the total number of constraints, formulation time and solver time when given identical input parameters.
As observed, the number of constraints in \textit{Opt B} is reduced by more than 90\%, which in turn decrease both formulation and solver time.
Additionally, it is noteworthy that the number of constraints with 4 directions is consistently lower than with 1 direction due to a reduction in the number of common nodes and edges among flights.}

\begin{table}[h]
\anon{\footnotesize}
\caption{\rev{Comparison of previous formulation Opt A and present formulation Opt B, in terms of number of constraints, formulation time and solver time. The decrease in the formulation and solver time is more than 90\% due to a massive decrease in the number of constraints}} \label{tab: optAvsoptB}
\resizebox{\textwidth}{!}
{%
\begin{tabularx}{\textwidth}{c*{8}{X}}
\toprule
\multirow{2}{*}{\textbf{\begin{tabular}[c]{@{}c@{}}Number of \\ Flights\end{tabular}}} & \multirow{2}{*}{\textbf{\begin{tabular}[c]{@{}c@{}}Number of \\ Directions\end{tabular}}} & \multicolumn{2}{c}{\textbf{Number of Constraints}}                        & \multicolumn{2}{c}{\textbf{Formulation Time (s)}}                             & \multicolumn{2}{c}{\textbf{Solver Time (s)}}                                  \\ \cmidrule(l){3-8} 
                                                                                       &                                                                                           & \multicolumn{1}{c}{\textbf{Opt A}} & \multicolumn{1}{c}{\textbf{Opt B}} & \multicolumn{1}{c}{\textbf{Opt A}} & \multicolumn{1}{c}{\textbf{Opt B}} & \multicolumn{1}{c}{\textbf{Opt A}} & \multicolumn{1}{c}{\textbf{Opt B}} \\ 
\multicolumn{1}{c}{}                                                                & \multicolumn{1}{c}{1}                                                                    & \multicolumn{1}{c}{1768}           & \multicolumn{1}{c}{458}            & \multicolumn{1}{c}{12}             & \multicolumn{1}{c}{1}              & \multicolumn{1}{c}{0}              & \multicolumn{1}{c}{0}              \\ 
\multicolumn{1}{c}{}                                                                & \multicolumn{1}{c}{2}                                                                    & \multicolumn{1}{c}{1647}           & \multicolumn{1}{c}{426}            & \multicolumn{1}{c}{1}              & \multicolumn{1}{c}{1}              & \multicolumn{1}{c}{0}              & \multicolumn{1}{c}{0}              \\ 
\multicolumn{1}{c}{5}                                                                & \multicolumn{1}{c}{3}                                                                    & \multicolumn{1}{c}{1646}           & \multicolumn{1}{c}{426}            & \multicolumn{1}{c}{2}              & \multicolumn{1}{c}{1}              & \multicolumn{1}{c}{0}              & \multicolumn{1}{c}{0}              \\ 
\multicolumn{1}{c}{}                                                                & \multicolumn{1}{c}{4}                                                                    & \multicolumn{1}{c}{1564}           & \multicolumn{1}{c}{402}            & \multicolumn{1}{c}{1}              & \multicolumn{1}{c}{1}              & \multicolumn{1}{c}{0}              & \multicolumn{1}{c}{0}              \\ \midrule
\multicolumn{1}{c}{}                                                               & \multicolumn{1}{c}{1}                                                                    & \multicolumn{1}{c}{8974}           & \multicolumn{1}{c}{1694}           & \multicolumn{1}{c}{7}              & \multicolumn{1}{c}{4}              & \multicolumn{1}{c}{2}              & \multicolumn{1}{c}{1}              \\ 
\multicolumn{1}{c}{}                                                               & \multicolumn{1}{c}{2}                                                                    & \multicolumn{1}{c}{7872}           & \multicolumn{1}{c}{1502}           & \multicolumn{1}{c}{6}              & \multicolumn{1}{c}{2}              & \multicolumn{1}{c}{2}              & \multicolumn{1}{c}{0}              \\ 
\multicolumn{1}{c}{10}                                                               & \multicolumn{1}{c}{3}                                                                    & \multicolumn{1}{c}{7602}           & \multicolumn{1}{c}{1438}           & \multicolumn{1}{c}{6}              & \multicolumn{1}{c}{4}              & \multicolumn{1}{c}{1}              & \multicolumn{1}{c}{0}              \\ 
\multicolumn{1}{c}{}                                                               & \multicolumn{1}{c}{4}                                                                    & \multicolumn{1}{c}{7622}           & \multicolumn{1}{c}{1438}           & \multicolumn{1}{c}{6}              & \multicolumn{1}{c}{2}              & \multicolumn{1}{c}{1}              & \multicolumn{1}{c}{0}              \\ \midrule
\multicolumn{1}{c}{}                                                               & \multicolumn{1}{c}{1}                                                                    & \multicolumn{1}{c}{26156}          & \multicolumn{1}{c}{3948}           & \multicolumn{1}{c}{27}             & \multicolumn{1}{c}{7}              & \multicolumn{1}{c}{6}              & \multicolumn{1}{c}{0}              \\ 
\multicolumn{1}{c}{}                                                               & \multicolumn{1}{c}{2}                                                                    & \multicolumn{1}{c}{22985}          & \multicolumn{1}{c}{3743}           & \multicolumn{1}{c}{26}             & \multicolumn{1}{c}{6}              & \multicolumn{1}{c}{6}              & \multicolumn{1}{c}{0}              \\ 
\multicolumn{1}{c}{15}                                                               & \multicolumn{1}{c}{3}                                                                    & \multicolumn{1}{c}{22006}          & \multicolumn{1}{c}{3356}           & \multicolumn{1}{c}{26}             & \multicolumn{1}{c}{5}              & \multicolumn{1}{c}{5}              & \multicolumn{1}{c}{0}              \\ 
\multicolumn{1}{c}{}                                                               & \multicolumn{1}{c}{4}                                                                    & \multicolumn{1}{c}{21662}          & \multicolumn{1}{c}{3276}           & \multicolumn{1}{c}{26}             & \multicolumn{1}{c}{5}              & \multicolumn{1}{c}{5}              & \multicolumn{1}{c}{0}              \\ \midrule
\multicolumn{1}{c}{}                                                               & \multicolumn{1}{c}{1}                                                                    & \multicolumn{1}{c}{56788}          & \multicolumn{1}{c}{7076}           & \multicolumn{1}{c}{112}            & \multicolumn{1}{c}{10}             & \multicolumn{1}{c}{380}            & \multicolumn{1}{c}{2}              \\ 
\multicolumn{1}{c}{}                                                               & \multicolumn{1}{c}{2}                                                                    & \multicolumn{1}{c}{55172}          & \multicolumn{1}{c}{6284}           & \multicolumn{1}{c}{159}            & \multicolumn{1}{c}{13}             & \multicolumn{1}{c}{705}            & \multicolumn{1}{c}{1}              \\ 
\multicolumn{1}{c}{20}                                                               & \multicolumn{1}{c}{3}                                                                    & \multicolumn{1}{c}{54780}          & \multicolumn{1}{c}{6188}           & \multicolumn{1}{c}{215}            & \multicolumn{1}{c}{12}             & \multicolumn{1}{c}{318}            & \multicolumn{1}{c}{1}              \\ 
\multicolumn{1}{c}{}                                                               & \multicolumn{1}{c}{4}                                                                    & \multicolumn{1}{c}{47739}          & \multicolumn{1}{c}{6050}           & \multicolumn{1}{c}{109}            & \multicolumn{1}{c}{8}              & \multicolumn{1}{c}{159}            & \multicolumn{1}{c}{0}              \\ \midrule
\multicolumn{1}{c}{}                                                               & \multicolumn{1}{c}{1}                                                                    & \multicolumn{1}{c}{395232}         & \multicolumn{1}{c}{29074}          & \multicolumn{1}{c}{6805}           & \multicolumn{1}{c}{38}             & \multicolumn{1}{c}{5285}           & \multicolumn{1}{c}{73}             \\ 
\multicolumn{1}{c}{}                                                               & \multicolumn{1}{c}{2}                                                                    & \multicolumn{1}{c}{342058}         & \multicolumn{1}{c}{25946}          & \multicolumn{1}{c}{6965}           & \multicolumn{1}{c}{42}             & \multicolumn{1}{c}{6972}           & \multicolumn{1}{c}{74}             \\ 
\multicolumn{1}{c}{40}                                                               & \multicolumn{1}{c}{3}                                                                    & \multicolumn{1}{c}{330040}         & \multicolumn{1}{c}{24826}          & \multicolumn{1}{c}{8812}           & \multicolumn{1}{c}{33}             & \multicolumn{1}{c}{3178}           & \multicolumn{1}{c}{72}             \\ 
\multicolumn{1}{c}{}                                                                                      & \multicolumn{1}{c}{4}                                              & \multicolumn{1}{c} {326420}                              & \multicolumn{1}{c}{24362}                               & \multicolumn{1}{c}{8874}                                & \multicolumn{1}{c}{32}                                  & \multicolumn{1}{c}{7283}                                & \multicolumn{1}{c}{78}                                  \\ \bottomrule
\end{tabularx} 
}
\end{table}
\newpage
\begin{table}[!h]
    \footnotesize
    \caption{Vertiminal Parameters}    \label{tab: vert symbols}
    \begin{tabular}{|l|m{0.95\textwidth}|}
        \hline
        $N_G$ & The set of gate nodes. Each physical gate node has an entrance node $n^{en}_g \in N^{en}_g$ and an exit node $n^{ex}_g \in N^{ex}_g\ \forall g \in N_G$. Each gate node has a holding capacity of $c_g$. \\
        \hline
        $N_R$ & The set of nodes representing all TLOF pads at the vertiminal. Each TLOF pad has an entrance node $n^{en}_r \in N^{en}_r$ and an exit node $n^{ex}_r \in N^{ex}_r\ \forall r \in N_R$. \\
        \hline
        $N^{n_r}_O$ & The set of nodes representing OFV boundary on a TLOF pad $n_r \in N_R$ \\
        \hline
        $N^{n_r}_F$ & The set of nodes representing the vertiexit reached from TLOF pad $ n_r \in N_R$ \\
        \hline
        \multicolumn{2}{|c|}{$N_O$ = $ \bigcup\limits_{n_r\in N_R} N^{n_r}_O$;\qquad $N_F$ = $\bigcup\limits_{n_r\in N_R} N^{n_r}_F$} \\
        \hline
        $N_T$ & The set of all nodes $n$ on the ground, where a node could be the intersection of two taxiways, an entrance or exit to a taxiway or a TLOF pad. \\
        \hline
        $N$ & $N_G \cup N_T \cup N_R \cup N_O \cup N_F$ \\
        \hline
        $L_G$ & Set of links connecting gate nodes $n_g \in N_G$ and taxiing nodes $n_t \in N_T$ \\
        \hline
        $L_T$ & Set of links connecting taxing nodes nodes $n_t \in N_T$ \\
        \hline
        $L_R$ & Set of links connecting taxing nodes $n_t \in N_T$ and TLOF pads $n_r \in N_R$. \\
        \hline
        $L_O$ & Set of links connecting TLOF pads $n_r \in N_R$ and OFV boundary $n_o \in N^{n_r}_O$ \\
        \hline
        $L_F$ & Set of links connecting OFV boundary nodes $n_o \in N^{n_r}_O$ and vertiexit nodes $n_f \in N^{n_r}_F\ \forall n_r \in N_R$ \\
        \hline
        $L$ & $L_T \cup L_R \cup L_O \cup L_F \cup L_G$ \\
        \hline
        $l_{a,b}$ & $ \in L$  A link connecting nodes $a$ and $b$ also represented as (a,b). \\
        \hline
        $G$ &$(N, L)$,  The vertiminal network \\
        \hline        
    \end{tabular}
\end{table}

\begin{table}[!h]
    \footnotesize
    \renewcommand{\arraystretch}{0.55}
    \caption{VTOL sets and parameters}    \label{tab: vtol sets}
    \begin{tabular}{|l|m{0.89\textwidth}|}
        \hline
         $A$ & The set of all VTOL. \\
        \hline
        $A^{Dep},A^{Arr}$ & The set of all departing and arriving VTOL respectively. \\
        \hline
        $A^{TAT}$ & $A^{Dep} \cap A^{Arr}$, the set of VTOL that have to turn around. \\
        \hline
        $A^n$ & The set of all VTOL whose route passes through node $n \in N$. \\
        \hline
        $\mathtt{ARAPPR}_i$ & $\forall i \in A^{Arr}$ the time at which an arriving flight $i$ enters the airspace of vertiminal i.e. enters vertiexit.\\
        \hline
        ${\mathtt{DRGATE}_i}$ & $\forall i \in A$ the time at which a departing flight $i$ is ready to leave the gate after passenger boarding.\\
        \hline
        $\gamma(i)$ & $ \ \forall i \in A$, a gate $g \in N_G$ assigned to VTOL $i$ \\
        \hline
        $\gamma^{en}(i)$ & Entrance node $n^{en}_g \in N^{en}_g$ of gate $g \in N_G$ assigned to VTOL $i \in A^{Arr}$ \\
        \hline
        $\gamma^{ex}(i)$ & Exit node $n^{ex}_g \in N^{ex}_g$ of gate $g \in N_G$ assigned to VTOL $i \in A^{Dep}$ \\
        \hline
        $\tau(i) $ & $ \forall i \in A$, a TLOF pad $n_r \in N_R$ assigned to VTOL $i$. $\tau^{Arr}(i)$ is a TLOF pad assigned to $i\in A^{Arr}$ while $\tau^{Dep}(i)$ is assigned to $i \in A^{Dep}$. \\
        \hline
        $\tau^{en}(i)$ & Entrance node $n^{en}_r \in N^{en}_r$ of the TLOF pad $n_r \in N_R$ assigned to VTOL $i \in A^{Dep}$. \\
        \hline
        $\tau^{ex}(i)$ & Exit node $n^{ex}_r \in N^{ex}_r$ of the TLOF pad $n_r \in N_R$ assigned to VTOL $i \in A^{Arr}$. \\
        \hline
        $\Lambda(i) $ & $ \forall i{\in }A$, a surface direction $l^i_{n_f,n_o} {\in} L_F , n_f \in N^{n_r}_F , n_o {\in} N^{n_r}_O$ assigned to VTOL $i$ from TLOF pad $n_r {\in} N_R$. $\Lambda^{Arr}(i)$ is the surface direction assigned to $i{\in} A^{Arr}$ while $\Lambda^{Dep}(i)$ is assigned to $i{\in} A^{Dep}$. \\
        \hline
        $\Lambda_F(i) $ & $ \forall i \in A$, a vertiexit node $n_f \in N_F$ on surface direction $\Lambda(i)$. \\
        \hline
        $\Lambda_O(i) $ & $ \forall i \in A$, an OFV boundary node $n_o \in N_O$ on surface direction $\Lambda(i)$. \\
        \hline
        $TAT_i$ & Turn Around Time of VTOL $i \in A^{TAT}$. \\
        \hline
        $TOT^{Arr}_i$ & TLOF pad Occupancy Time of arriving VTOL $i$. It is the time a VTOL takes to stop its motor and cool down before it can exit the TLOF pad. \\
        \hline
        $TOT^{Dep}_i$ & TLOF pad Occupancy Time of departing VTOL $i$. It is the time a VTOL takes to reach the centre of the TLOF pad, start its motor, and prepare for take-off. \\
        \hline
        $W^{tsep}_{ij}$ & Required safe separation time at landing or takeoff of VTOL $i$ from its immediate trailing VTOL $j$, $\forall i,j \in A, i \neq j$. \\
        \hline
    \end{tabular}
\end{table}

\begin{table}[!h]
    \centering
    \renewcommand{\arraystretch}{1.5}
    \caption{Description of Weight Symbols}    \label{tab:weight_symbols}
    \begin{tabular}{|l|m{0.95\textwidth}|}
        \hline
        ${W_g}$ & Weight assigned to a departing VTOL for the time it expends at the gate. $0 \leq {W_g} \leq 1$. \\
        \hline
        ${W_t^{Dep}}$ & Weight assigned to a departing VTOL for the taxiing time it expends. The requirement ${W_g} \leq {W_t^{Dep}} \leq 1$ ensures that a VTOL should spend more time on the gate rather than on taxiways so as to avoid congestion and save energy. Also, it is beneficial for a multirotor to fly (in the case of air taxiing) at the maximum optimal speed in order to reduce energy consumption~\cite{RRJTOSN}.
        \\
        \hline
        $W^{Dep}_r$ & Weight assigned to a VTOL for the time it spends on the TLOF pad and inside the OFV during take-off. TLOF pad and OFV are highly constrained resources and a bottleneck; hence, this weight value should be kept high. $0 \leq W^{Dep}_r \leq 1$. \\
        \hline
        $W^{Arr}_r$ & Weight assigned to a VTOL for the time it spends inside the OFV and on the TLOF pad during landing. TLOF pad and OFV are highly constrained resources and a bottleneck; hence, this weight value should be kept high. $0 \leq W^{Arr}_r \leq 1$. \\
        \hline
        ${W_t^{Arr}}$ & Weight assigned to an arriving VTOL for the taxiing time it expends. The requirement $0 \leq {W_t^{Arr}} \leq W_r$ ensures that the VTOL will leave the TLOF pad at the earliest. \\
        \hline
        ${W_c^{Dep}}$ & Weight assigned to a departing VTOL to the time it expends to reach vertiexit from a OFV boundary. $0 \leq {W_c^{Dep}} \leq 1$. \\
        \hline
        ${W_c^{Arr}}$ & Weight assigned to an arriving VTOL to the time it expends to reach the OFV boundary from a vertiexit. $0 \leq {W_c^{Arr}} \leq 1$. \\
        \hline
        $W_q$ & Weight assigned to a VTOL for spending time on holding space of a gate after arrival. $0 \leq W_q \leq 1$. \\
        \hline
    \end{tabular}
\end{table}


\section{Throughput Capacity Equations}
\label{sec: tc}
\renewcommand{\descp}[1]{\textcolor{purple}{#1}}
\hidetxt{This section explores the analysis of the \verti 's throughput capacity.
Similar to the traditional airports in Air Traffic Flow Management (ATFM),
The throughput capacity of a \verti\ defines the extent to which the UTM can accommodate and manage urban air traffic, thus critically affecting UTM's overall performance.}

\textcolor{black}{
In the previous section, our objective was to minimize a weighted sum of delays, subject to constraints arising from vertiminal operations. While the weighted sum of delays is optimized, we do not know if the optimal strategy in Section~\ref{sec: MILP} ends up affecting the throughput of the vertiminal. To answer this question, we study the maximum throughput that can be achieved -- this is what we call ``throughput capacity.''}
\hidetxt{The maximum throughput of a conventional airport is typically determined by the capacity of its airfield, with runway capacity being a critical determinant.
Thus, similarly, }
The capacity of the system comprising of: \reva{(a) TLOF pads, (b) taxiways and (c) gates at the apron} \streva{(a) gates at the apron, (b) taxiways and (c) TLOF pads} would determine the maximum throughput capacity of a \verti.
We analyze each of these elements independently \hidetxt{of others using the} and base our analysis 
on \hidetxt{``Airport Systems Planning, Design, and Management" }\cite{de2020airport}. 
{ Finally, the throughput obtained using the optimal strategy in Section~\ref{sec: MILP} is compared with the throughput capacity.
}
\subsection{TLOF pad system}
\label{sec: tc TLOF}

\begin{figure}[h]
    \centering
    \disableFig
    {\includegraphics[width=0.5\linewidth]{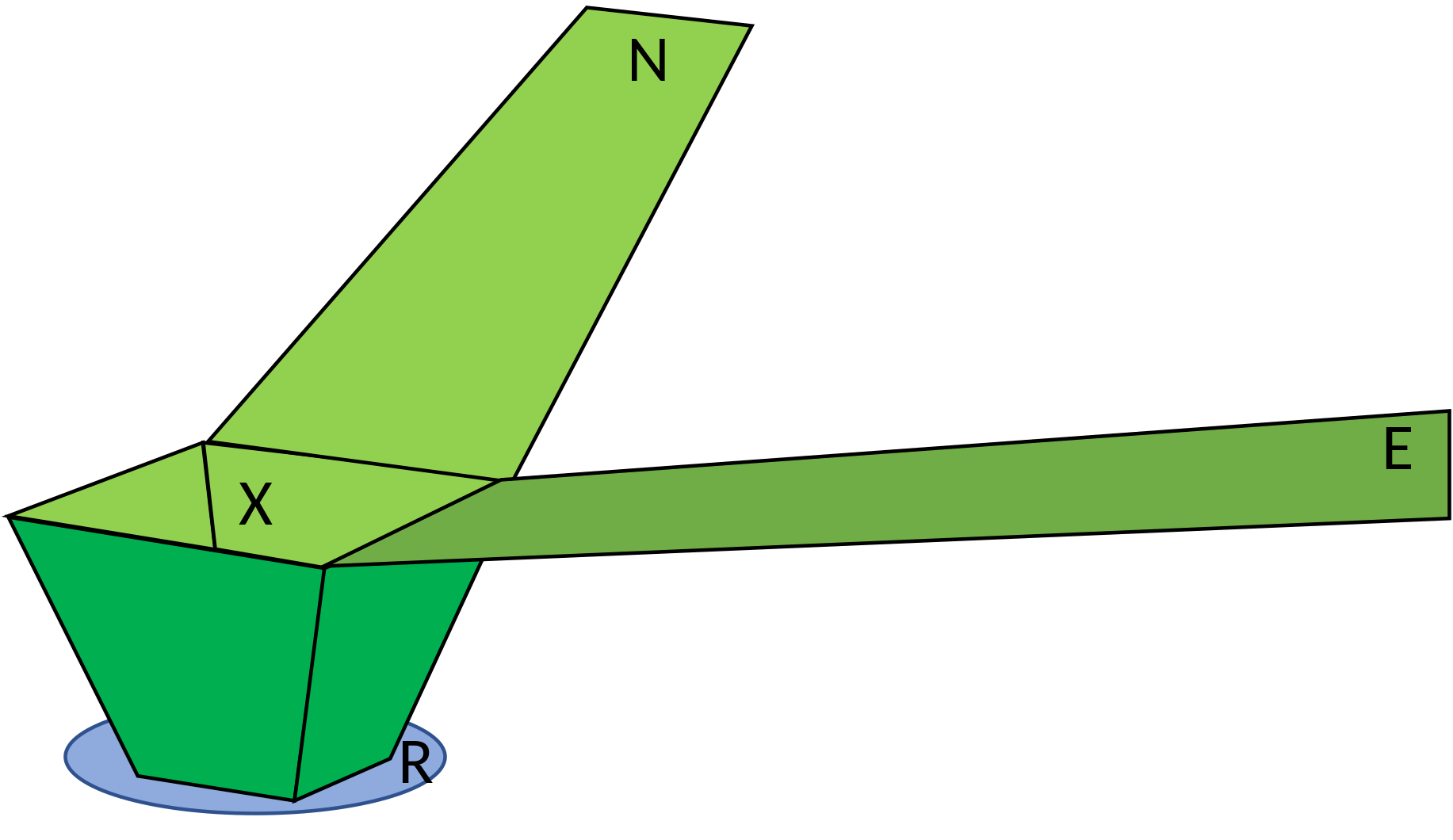}}
    \caption{TLOF pad system with two surface directions `X-N' and `X-E'}
    \label{fig: TLOF capacity}
\end{figure}

The TLOF pad system consists of TLOF pad, OFV and surface directions as illustrated in Figure~\ref{fig: TLOF capacity}.
We analyse the throughput capacity by considering the arrivals and departures of VTOLs to and from the system.
While a few factors such as visibility, precipitation, wind direction, etc. have been ignored, the following factors are considered in the capacity calculations: 
\begin{itemize}
    \item Number of TLOF pads
    \item Number of surface directions
    \item Separation requirements between VTOLs imposed by the ATM system
    \item Mix of VTOL classes using the \verti
    \item Mix of movements such as arrivals, departures, or mixed on each TLOF pad and their sequencing
\end{itemize}
\hidetxt{
Further factors that may be regarded but are not in the scope and left for future study
\begin{itemize}
    \item Visibility, cloud ceiling, and precipitation
    \item Wind direction and strength
    \item State and performance of the UTM system
    \item Noise-related and other environmental considerations and constraints
\end{itemize}
}
We define the term \textit{maximum throughput capacity} as the maximum number of movements that can be performed in unit time on a TLOF pad without violating ATM rules.\strev{ such as wake vortex and other separation timing between two VTOLs.}
We first calculate for a single TLOF pad and then extend our calculations to multiple TLOF pads. 
On a TLOF pad, there are 4 possible sequences of movement pairs: Arrival-Arrival $(\mathbb{A}\mathbb{A})$, Departure-Departure $(\mathbb{D}\mathbb{D})$, Arrival-Departure $(\mathbb{A}\mathbb{D})$ and Departure-Arrival $(\mathbb{D}\mathbb{A})$.
Recall that ATM regulations require safe separation distance on surface directions and wake vortex time separation on the TLOF pad. 
These separation requirement values depend on the classes of the VTOLs involved.
Additionally, during departure or arrival operations, only 1 VTOL can occupy the OFV.
The throughput capacity is determined by the time separation enforced by ATM rules for various VTOL classes executing different movements. 

The different time parameters we are going to utilise for analysis are:
\begin{enumerate}
    \item $t^{sep}_{ij}$: The safe separation distance required on the surface direction between VTOL $i$ and $j$, $i$ followed by $j$, is converted into time separation (constraint~\eqref{TaxiSep and climbSep C20}).
    \item $t^w_{ij}$: Wake vortex separation on TLOF pad between VTOL $i$ and $j$
    \item $t^{TOT}_i$: Occupancy time of a VTOL $i$ entering a TLOF pad before take-off or after landing on a TLOF pad before exit.
    \item $t^{R-X}_i, t^{X-R}_i$: OFV travel time by a VTOL during departure or arrival, respectively.
    \item $t^{X-N}_i, t^{N-X}_i$: Surface direction travel time by a VTOL during departure or arrival, respectively.
\end{enumerate}
The following equations calculate the maximum time $T^{\mathbb{M}_1\mathbb{M}_2}_{ij}$ required in different sequences of movements $(\mathbb{M}_1\mathbb{M}_2),where\ \mathbb{M}_1,\mathbb{M}_2 \in \{ \mathbb{D}, \mathbb{A}\}$, and VTOL $i$ is followed by VTOL $j$.

\begin{itemize}
    \item Arrival-Arrival: Two VTOLs are arriving from vertiexit to the TLOF pad. If the VTOLs $i$ and $j$ are arriving from different surface directions, then set $t^{sep}_{ij}$ to 0.
        \begin{equation}
            T^{\mathbb{A}\mathbb{A}}_{ij} = max(t^{sep}_{ij},t^w_{ij}, t^{X-R}_i+t^{TOT}_i)
            \label{eq tAA}
        \end{equation}
    \item Departure-Departure: Two VTOLs are departing from the TLOF pad to vertiexit. If the VTOLs $i$ and $j$ are departing in different surface directions, then set $t^{sep}_{ij}$ to 0.
        \begin{equation}
            T^{\mathbb{D}\mathbb{D}}_{ij} = max(t^{sep}_{ij},t^w_{ij}, t^{TOT}_i+t^{R-X}_i)
            \label{eq tDD}
        \end{equation}
    \item Arrival-Departure: When an arrival is followed by a departure on a TLOF pad. In case the VTOLs $i$ and $j$ are using different surface directions, set $t^{N-X}_i$ to 0.
        \begin{equation}
            T^{\mathbb{A}\mathbb{D}}_{ij} = max(t^{N-X}_i+t^{X-R}_i+t^{TOT}_i,t^w_{ij})
            \label{eq tad}
        \end{equation}
    \item Departure-Arrival: When a departure is followed by an arrival on a TLOF pad. In case the VTOLs $i$ and $j$ are using different surface directions, set $t^{X-N}_i$ to 0.
        \begin{equation}
            T^{\mathbb{D}\mathbb{A}}_{ij} = max(t^{TOT}_i+t^{R-X}_i+t^{X-N}_i,t^w_{ij})
            \label{eq tda}
        \end{equation}
\end{itemize}

\hidetxt{For any movement pair $(\mathbb{M}_1\mathbb{M}_2)$, the maximum throughput can be calculated by considering the VTOL classes that give the minimum time of the movement.}
The maximum throughput of the movement pair $\mathbb{M}_1\mathbb{M}_2$ per unit time is given as 
\begin{equation}
    \mu^{\mathbb{M}_1\mathbb{M}_2} =  \frac{1}{min(T^{\mathbb{M}_1\mathbb{M}_2}_{ij})} , \mathbb{M}_1,\mathbb{M}_2 \in \{ \mathbb{D}, \mathbb{A}\}, i,j \in A, i \neq j
    \label{eq movement capacity}
\end{equation}
The work in \hidetxt{ analysis in the book ``Airport Systems Planning, design, and Management" }\cite{de2020airport} calculates the expected capacity of a runway by considering all the possible aircraft classes and all possible permissible pairs of movements that have occurred in the past and then using a probability of occurrence for each of these events. 
Due to lack of any real-world data on UTM, \hidetxt{we'll be calculating only} our calculation is limited to the maximum throughput of any system. 
The maximum throughput of any movement pair $(\mathbb{M}_1\mathbb{M}_2)$ is calculated by evaluating the VTOL classes that have the minimum values of the time parameters. 
\strev{In the random sequence of arrivals and departures, the repeated occurrence of the movement pair $(\mathbb{D}\mathbb{A})\ or\ (\mathbb{A}\mathbb{D})$ has a low probability.
We take the maximum time of the movement pairs $(\mathbb{A}\mathbb{A})\ and\ (\mathbb{D}\mathbb{D})$ to find the bottleneck in the system and define the capacity.}
\rev{We take the minimum time of the all movement pairs to find the bottleneck in the system and define the throughput capacity.}
\begin{equation}
    T_{ij} = min(T^{\mathbb{A}\mathbb{A}}_{ij}, T^{\mathbb{D}\mathbb{D}}_{ij}, T^{\mathbb{A}\mathbb{D}}_{ij}, T^{\mathbb{D}\mathbb{A}}_{ij}) 
    \label{eq verti capacity}
\end{equation}

The maximum throughput capacity of the TLOF pad system, per unit time, is then given as: 
\begin{equation}
\mu^{TLOF} =  \frac{1}{T_{ij}} 
\label{eq verti thru}
\end{equation}

A TLOF pad can facilitate both arrival and departure movements or can also cater\hidetxt{exclusively} to a single type of movement. 
In the case of a vertiminal equipped with multiple TLOF pads, its capacity is decided according to the configuration, i.e., there can be several combinations of assigning arrivals and departure movements to the TLOF pads. 
For instance, in a vertiminal with two TLOF pads, one configuration might utilize both pads for both arrival and departure. In contrast, another configuration might designate one TLOF pad exclusively for arrivals and the other for departures. 
In either case, the overall capacity of the vertiminal is determined by the combined throughput capacity of each individual TLOF pad.
\begin{equation}
    \mu^{TLOF}_c = \sum_{r \in N_R}  \frac{1}{T^{r, \rev{c}}_{ij}} , \rev{\qquad r \in N_R,} c \in C
    \label{eq multi verti cap}
\end{equation}
where $T^{r,c}_{ij}$ is calculated for each TLOF pad $r$ \strev{$r \in N_R$} using~\eqref{eq verti thru} \rev{for a configuration ($c$) from a} set of configurations.

\subsection{Taxiway system}
\label{sec: tc taxi}

\begin{figure}[h]
    \centering
    \disableFig
    {\includegraphics[width=0.6\linewidth]{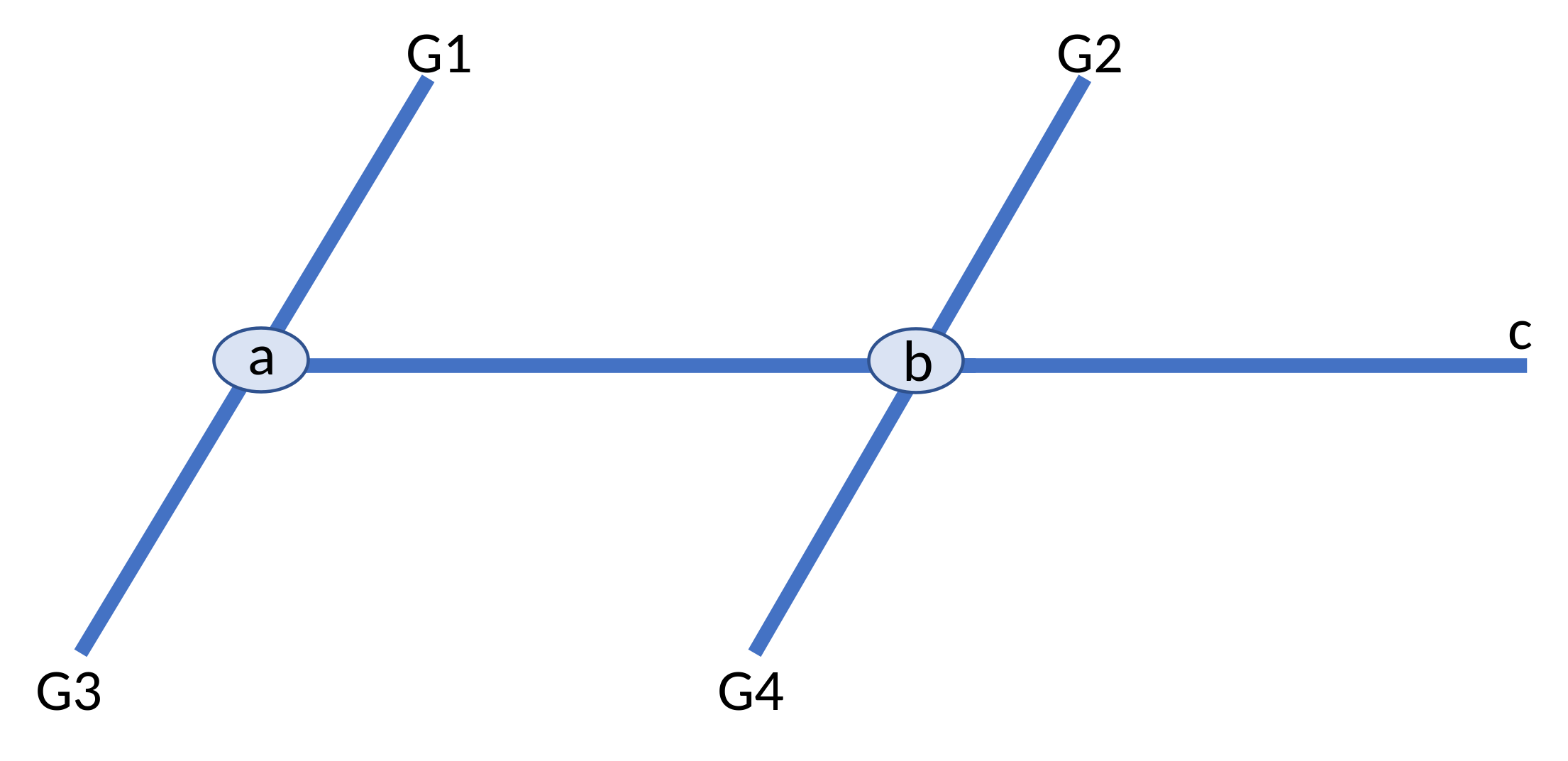}}
    \caption{Taxiway system connecting TLOF pad and gates}
    \label{fig: taxi capacity}
\end{figure}

The traditional airport always has a taxiway running parallel to the runway.
These long, well-designed taxiways are not a factor limiting the capacity of an airport. 
However, the taxiway system of \verti\ would be different as they are shorter,
linking TLOF pads' entry and exit points to the gate exit and entry points, respectively, as shown in Figure~\ref{fig: taxi capacity}.
There might be a single taxiway link or multiple taxiways with intersections among them on the \verti, each being a half-duplex link.
Considering the gates and TLOF pads as sources and sinks, respectively (or even vice versa), the taxiway system can be conceptualized as a flow network problem.
By deriving each link's flow capacity, the flow network's maximum capacity can be calculated using the max-flow min-cut theorem \cite{mfmc}.

Each taxiway link has a static and dynamic capacity.
Static capacity can be defined as the maximum number of stationary vehicles on it, while the dynamic capacity is the maximum number of vehicles passing through it per unit time, which is independent of taxiway length.
Assuming uniform taxi velocities for all VTOL classes, the minimum sum of vehicle length and safe separation distance between two vehicles determines the dynamic capacity on any taxiway link and, thus, its flow rate. 
Given the size of the VTOL vehicle as $d_{len}$, the separation distance $d_{sep}$ and the maximum taxi-velocity of the VTOLs as $v_{taxi}$, the maximum flow rate of a taxiway link can be calculated as: 
\begin{equation}
    maxFlowRate =  \frac{1}{(d_{len}+d_{sep})/v_{taxi}} 
    \label{eq taxi flow}
\end{equation}
If the three parameters in~\eqref{eq taxi flow} remain constant for each taxiway link, the maximum flow rate remains uniform across the entire taxiway network. 
However, varying $d_{sep}$ and $v_{taxi}$ for each link necessitates the use of the max-flow min-cut theorem to calculate the overall maximum flow rate for the network($\mu^{taxi}$).

\subsection{Gate system}
\label{sec: tc gate}

\begin{figure}[h]
    \centering
    \disableFig
    {\includegraphics[width=0.5\linewidth]{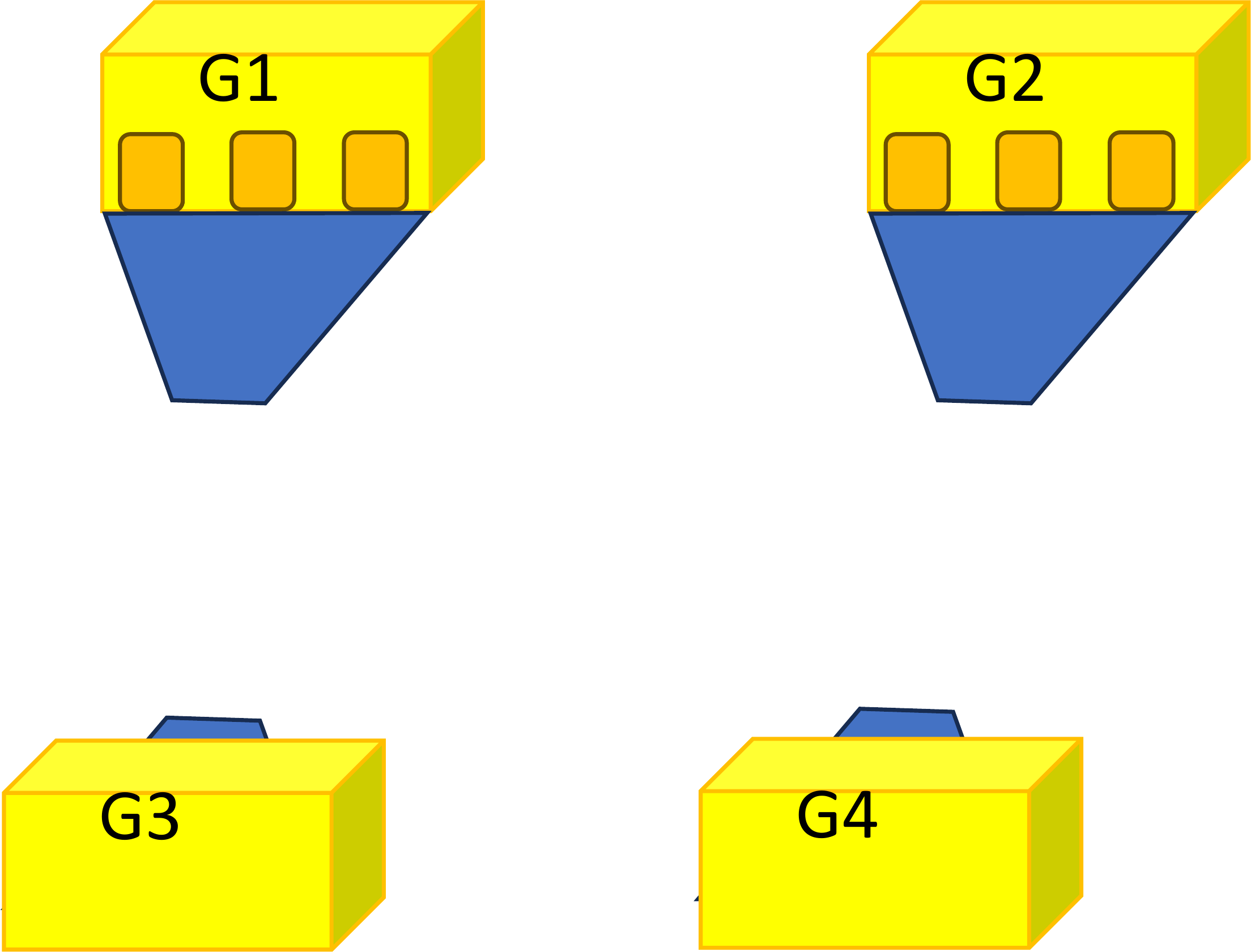}}
    \caption{Gate system with 4 gates, each having 3 slots}
    \label{fig: gate capacity}
\end{figure}
The gate system on a \verti\ can consist of multiple gates, each having multiple parking slots for VTOLs, as shown in Figure~\ref{fig: gate capacity}. 
The most basic and standard way to calculate the capacity of a gate system is the total number of VTOLs that can stand on a \verti\ simultaneously, referred to as static capacity ($\mathbb{C}_G = \sum_{g \in N_G} c_g$).
However, a better measure of the capacity is the dynamic capacity, defined as the number of VTOLs per unit time that can be accommodated by the gate system and this is affected by the VTOLs' turnaround times. 
For calculating the \textbf{maximum} throughput of the gate system, the class of vehicle having the minimum turnaround time $TAT$ should be considered.
The maximum throughput is then calculated as: 
\begin{equation}
    \mu^G =  \frac{\mathbb{C}_G}{TAT_{small}} 
    \label{eq tc gate}
\end{equation}
$\mu^G$ is an ideal value assuming immediate slot refilling.

\subsection{Vertiminal Throughput}
Till now, we have shown individual throughput calculations for TLOF pad, taxiway and gate systems, considering them independently. 
The overall \verti 's throughput is limited by the minimum of the three systems as shown 
\begin{equation}
    \mu^{\verti} = \min(\mu^{TLOF},\mu^{taxi},\mu^{G})
    \label{eq tc verti}
\end{equation}

\section{Evaluation of the results}
\label{sec: results}
We used MATLAB version 2022a (9.12) with an optimisation toolbox of version 9.3 on a Desktop PC equipped with an i7-11700F processor having 16 cores (8 physical) @ 2.5GHz and 64GB RAM to solve the optimisation problem described in \ref{sec: MILP}. Our implementation is available on GitHub for reproducibility \cite{github}.

\begin{figure}[h!]
  \centering
  \disableFig
  {\includegraphics[width=\linewidth]{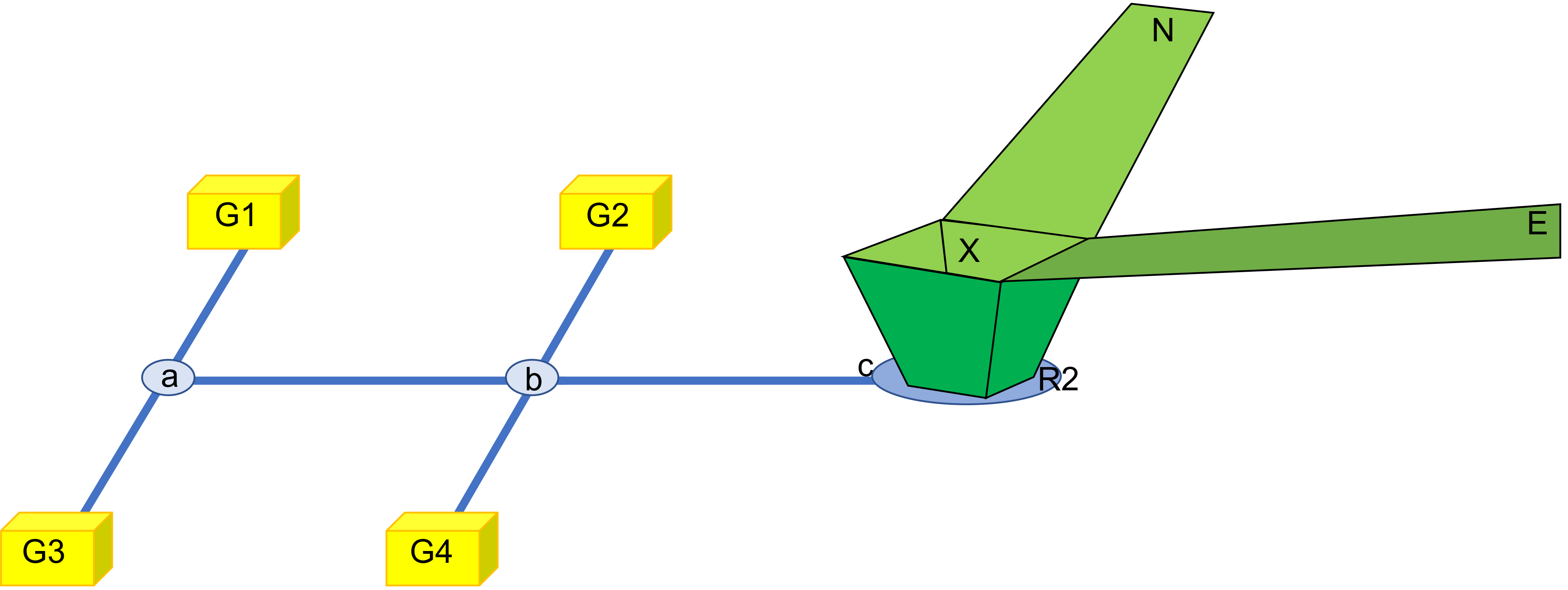}}
  \caption{Sample topology with 4 gates, 3 parking slots each and 2 surface directions.}
  \label{fig:topology}
\end{figure}

Figure~\ref{fig:topology} shows the topology used in our computational setup. 
All taxi edges in the network are equal in length, and similarly, all surface direction edges are equal. However, the length of the surface direction edges is greater than that of the taxi edges.
\strevcnf{We consider five categories of VTOLs, assuming that the taxiing and climbing speeds are the same for all UAM classes.
However, distance separation and wake vortex time requirements vary among VTOL classes.}
Table~\ref{tab: param} mentions the weights used in objective~\eqref{obj  eqn} of MILP and other parameters that encompass several edge lengths and speeds. 
We randomly assign gate, surface direction, $DRGATE_i$ and $ARAPPR_i$ to VTOLs.
The large constant $M$ is set as $(\lceil (|A|/10) \rceil +1)*2000$.

\begin{table}[h]
    \centering
    \caption{Weights and Parameters used in the MILP (* Used in Section~\ref{sec: throughput results})}
    \label{tab: param}
    \begin{tabular}{c c}
        \begin{minipage}{0.48\textwidth}
            \centering
            \begin{tabular}{|cc|}
                \hline
                $W_g$ & 0.2 \\
                \hline
                $W^{Dep}_t$ & 0.8 \\
                \hline
                $W^{Dep}_r$ & 1 \\
                \hline
                $W^{Dep}_c$ & 0.7 \\
                \hline
                $W_Q$ * & 0.1 \\
                \hline
                $W^{Arr}_c$ * & 0.7 \\
                \hline
                $W^{Arr}_r$ * & 1 \\
                \hline
                $W^{Arr}_t$ * & 0.8 \\
                \hline
            \end{tabular}
        \end{minipage}
        &
        \begin{minipage}{0.48\textwidth}
            \centering
            \begin{tabular}{|cc|}
                \hline
                Gate taxiway length & 30 units \\
                \hline
                Taxiway edge length & 45 units \\
                \hline
                OFV length & 75 units \\
                \hline
                Surface direction length & 300 \\
                \hline
                Average Taxi-speed & 6 units/sec \\
                \hline
                Max OFV climb speed & 17.14 units/sec \\
                \hline
                Max Surface direction speed & 23.73 units/sec \\
                \hline
                Separation on taxi (Small) & 5 units \\
                \hline
            \end{tabular}
        \end{minipage}
    \end{tabular}
\end{table}

\subsection{Results from the MILP formulation for VTOL departures}
\label{sec: MILP results}
 
 The solution provides the exact schedule of a VTOL movement from gate to vertiexit --- the times at which a VTOL: \hidetxt{is expect to} (i) departs from the gate, (ii) crosses each node on in its path, (iv) arrives at the TLOF pad, (v) crosses OFV boundary and (vi) exits from the vertiexit.
 
The $MinTravelTime$ for a VTOL is calculated as $t^{\Lambda^{Dep}_F(i)}_i - \mathtt{DRGATE}_i, i\in A^{Dep}$, where $\mathtt{DRGATE}_i$ is the desired time to depart from the gate and $t^{\Lambda^{Dep}_F(i)}_i$ is time to reach the vertiexit. It is achievable only when no other VTOLs are present in the vertiminal.
By comparing the actual time (provided by the MILP solution) with $MinTravelTime$, we determine the \textit{ExcessDelay}\hidetxt{ over VTOL's route}.
Our\hidetxt{initial} goal is to observe the effect of multiple surface directions on \textit{ExcessDelay} incurred by departing flights. 
We consider a total of $4$ surface directions from the OFV boundary (`X' in Figure~\ref{fig:topology}). 
The total number of runs is 24 (6 different sets of flights and 4 directions).
We compare the \textit{ExcessDelay} obtained from our formulation with that of FCFS scheduling.
Our implementation of FCFS uses the same set of constraints described in Section~\ref{sec: MILP} with an additional constraint that forces the VTOLs to take off according to the order of their $\mathtt{DRGATE}_i$.
We recall that a route is defined by the following $4$ elements: departing gate, a taxiing path from gate to TLOF pad, TLOF pad and surface direction.
For comparison, in each run MILP and FCFS are provided with the same set of flights along with their route and $\mathtt{DRGATE}_i$.

\begin{figure}[h]
    \hspace{-35pt}
    \disableFig
    {\includegraphics[width=1.1\linewidth]{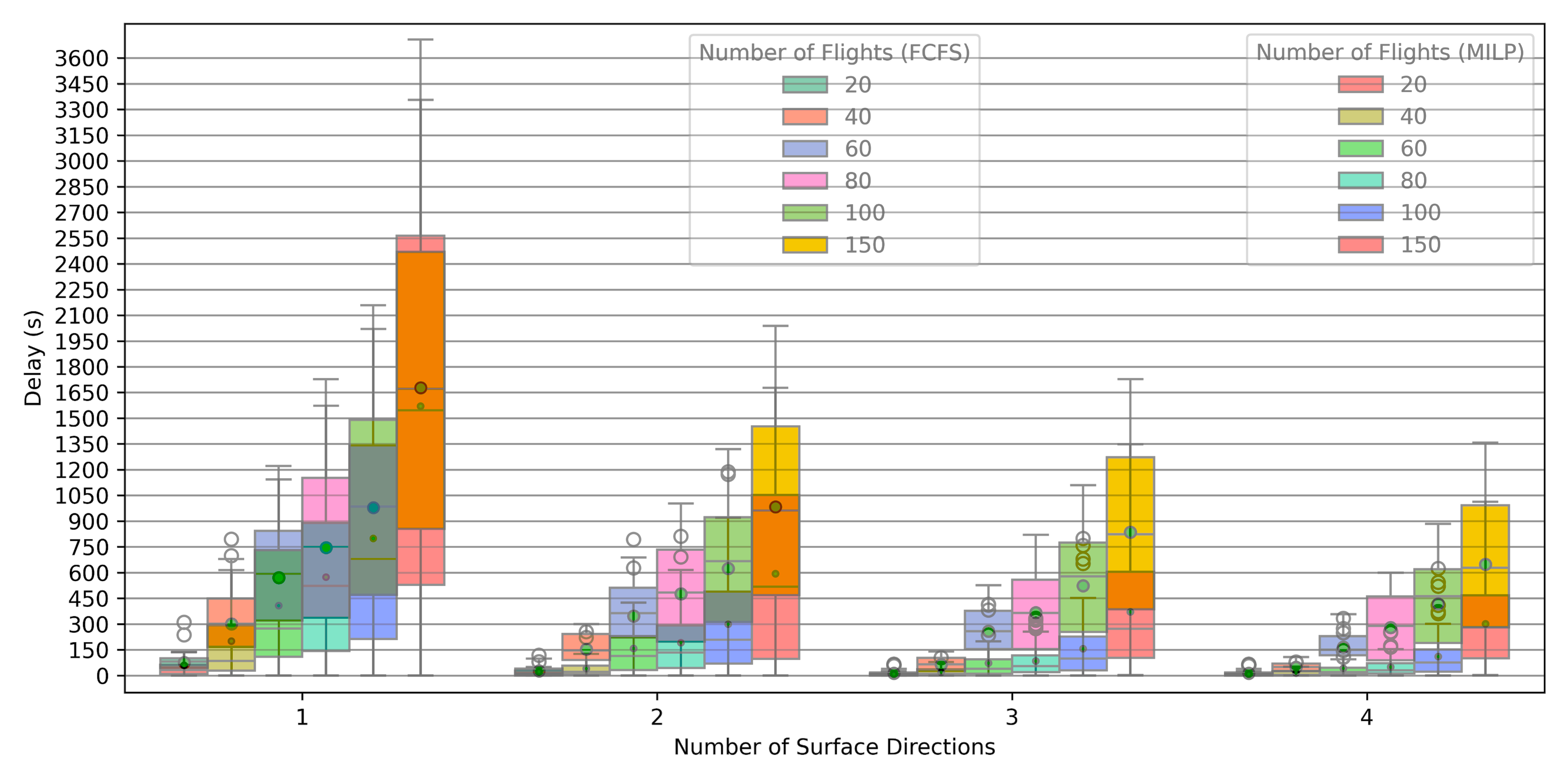}}
    \caption{Results from MILP overlapped with FCFS for multiple surface directions. The results are generated for 20, 40, 60, 80, 100, and 150 flights (indicated by different colours). It shows a reduction in delay by increasing the number of surface directions. The mean value of each box plot is marked in a small circle for MILP and a big circle for FCFS. Observe that the FCFS mean delay is aligned with its median and is always higher than the mean delay of MILP.}
    \label{fig:MILPvsFCFS}
\end{figure}

\hidetxt{In the departure operation, } When two successive VTOLs use \hidetxt{taking} different surface directions,\hidetxt{after the take-off} since there is no separation requirement it is sufficient for the follower to take off by adhering \hidetxt{ the immediate follower can take off by adhering }only to wake vortex separation, thus reducing its overall delay i.e. \textit{ExcessDelay}. 
This delay reduction is observed in Figure~\ref{fig:MILPvsFCFS} by comparing the delay distribution among flights with an increase in the number of surface directions.
From Figure~\ref{fig:MILPvsFCFS}, we see a delay reduction of about 50\% as we increase the number of surface directions from 1 to 2.
Furthermore, as we increase the number of flights, the delay increases due to congestion on \verti.
An analysis of the delay distribution for a flight set between our MILP and FCFS scheduling reveals notable differences. 
The third quartile (75\%) of MILP is lower than the third quartile of FCFS (with the exception of 
\hidetxt{Except in the case of} single surface direction and 150 flights). 
\hidetxt{However, }The mean and median of MILP are always less than that of FCFS.
Our MILP delays tend to skew away from the maximum delay, as evidenced by a higher mean value than the median. 
In contrast, the FCFS delays are uniformly distributed among flights, with coinciding mean and median values. 
In the presence of multiple surface directions, our formulation provides a significantly lower mean delay. 
More than 50\% of the flights have a 50\% reduction in delay. 
In summary, \hidetxt{This suggests that }our MILP formulation prioritises minimising overall flight delays. 

\begin{figure}[!h]
    \centering
    \begin{subfigure}[b]{0.32\textwidth}
        \centering
        \disableFig
        {\includegraphics[width=\textwidth]{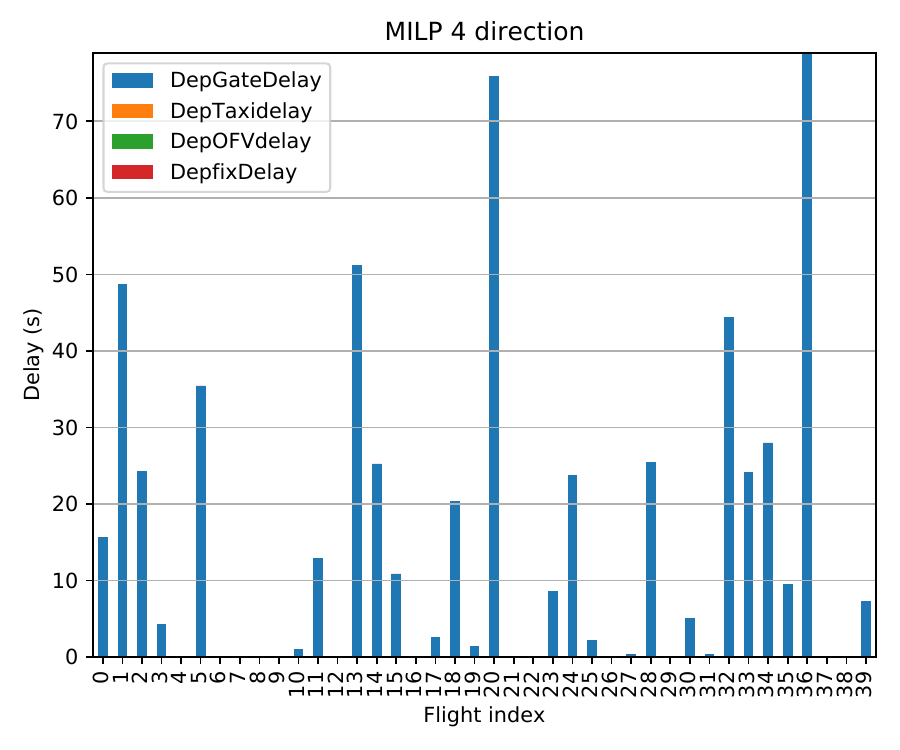}}
        \caption{MILP}
        \label{fig:MILP4}
    \end{subfigure}
    \hfill
    \begin{subfigure}[b]{0.32\textwidth}
        \centering
        \disableFig
        {\includegraphics[width=\textwidth]{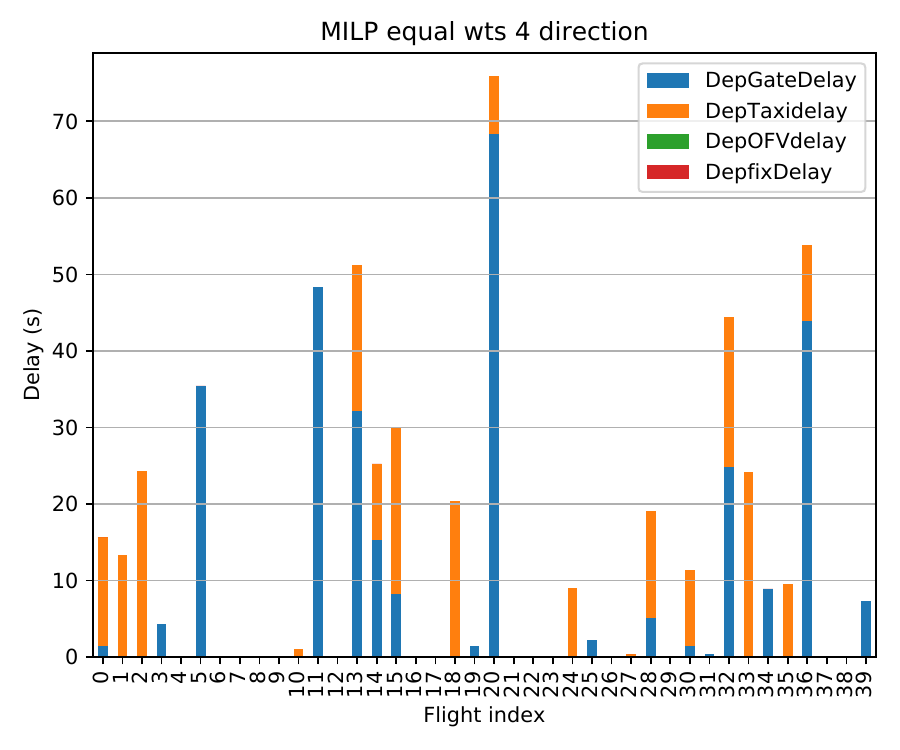}}
        \caption{MILP with equal weights}
        \label{fig:MILP4eq}
    \end{subfigure}
    \hfill
    \begin{subfigure}[b]{0.32\textwidth}
        \centering
        \disableFig
        {\includegraphics[width=\textwidth]{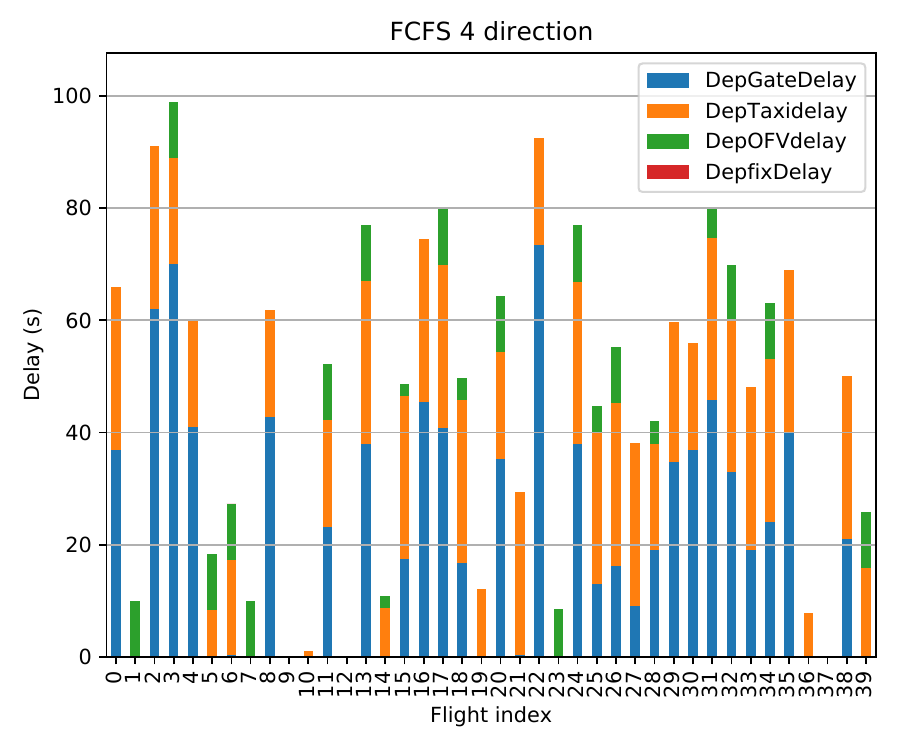}}
        \caption{FCFS}
        \label{fig:FCFS4}
    \end{subfigure}
    \caption{Distribution of delays of 40 flights among gate delay (blue bars), taxing delay (orange bars), OFV delay (green bars) and climb delay on 4 directions. For MILP the weights used are $W_g=0.2, W^{Dep}_t=0.8, W^{Dep}_r=1, W^{Dep}_c=0.7$. It can be observed that the MILP with equal weights faces taxing delays while FCFS additionally faces OFV delays, causing delays on other flights.}
    \label{fig:barplots}
\end{figure}

The bar plots shown in Figure~\ref{fig:barplots} represent the distribution of \textit{ExcessDelay} components such as gate delay (waiting at the gate), taxiing delay, OFV delay, and climb delay for each flight. 
Taxing delays signify congestion on taxiways, while OFV delay indicates a lack of availability of resources such as TLOF pad and OFV.\hidetxt{ to other VTOLs, thus causing further delay.}
Due to congestion on taxiways, a VTOL may not be able to fly or move at the optimal speed that may result in energy wastage \cite{RRJTOSN}. 
Thus, taxi delays are undesirable and it is more energy-efficient to wait at the gate.
\hidetxt{ For the computation, whose results are shown in the bar plots, }
We used 40 flights with 4 surface directions and compared the results with FCFS scheduling. 
Figure~\ref{fig:MILP4} shows the MILP formulation where \hidetxt{the delay} all the flights experience only gate delay. 
Additionally, we analyse the impact of equal weights used in the objective \hidetxt{of MILP }to study the VTOL delay. 
This effectively changes \hidetxt{Equal weights effectively change }the objective to minimise the sum of delay 
i.e.,  $\sum_{i \in A^{Dep}} (t^{\Lambda^{Dep}_F(i)}_{i} - \mathtt{DRGATE}_i)$.
Figure~\ref{fig:MILP4eq} shows the MILP with equal weights. The delay constitutes gate delay and taxiing delay. The taxiing delay (orange bars) occurs as flights are slowly moving on taxiways instead of waiting at the gate, even though the overall delay is at par with MILP (shown in Figure~\ref{fig:MILP4}).
Figure~\ref{fig:FCFS4} shows gate delay, taxiing delay and OFV delay (green bars) for FCFS scheduling. The OFV delay component is large compared to the zero OFV delay in MILP. 
In summary, the delay breakup not only demonstrates the importance of weighted delays in the objective function, it also shows that FCFS scheduling leads to increased delay incurred by the VTOLs.

\subsection{Results from Throughput Capacity}
\label{sec: throughput results}
Recall that in  \formatone{Section~\ref{sec: tc}}\formattwo{last three subsections}, we discussed the capacity equations for: (a) TLOF pad system, (b) Taxiway system, and (c) Gate system. 
We determine the maximum achievable throughput of the \verti\ using the equations and the parameters shown in Table~\ref{tab: param}.
We had considered the smallest class of VTOLs because they have the shortest turn-around time and minimum separation distance requirements.
\formattwo{We will use the same system settings, topology (Figure~\ref{fig:topology}) and parameters (Table~\ref{tab: param}) as used in Section \ref{sec: MILP results} for computation.}
Using the time stamp of the VTOLs to reach the TLOF pad\hidetxt{ or gate}, we analyze the throughput achieved by the \rev{MILP explained}\strev{optimal strategy for solving the problem} in 
Sec~\ref{sec: MILP}.
We have varied the number of flights and the number of surface directions for all possible combinations of movements. 
For the computation of the TLOF pad system, we will use only $A^{Arr}$ (or $A^{Dep}$) where the vertiexit is a source (sink), and the gates act as sinks (sources).

\hidetxt{We determine the maximum achievable throughput of the \verti\ using the equations in Section~\ref{sec: tc} and the parameters shown in Table~\ref{tab: param}.}
The following are the two sets of  parameters used in comparison of throughput analysis using MILP and the equations from Section~\ref{sec: tc}:
\begin{itemize}

    \item Set1: Separation requirement on surface direction is 75 units, Turn Around Time (TAT) on the gate is 90 seconds.
    \item Set2: Separation requirement on surface direction is 280 units, Turn Around Time (TAT) on the gate is 120 seconds.
\end{itemize}
\subsubsection{TLOF pad system}\label{Sec: TLOF pad sys results}
\begin{table}[h]
    \centering
    \caption{TLOF pad system time parameters}    \label{tab: tlof system}
    \begin{tabular}{lcc}
    \toprule
    \textbf{\reva{Short description}} & \textbf{Symbol} & \textbf{Value} \\ \hline
    Separation requirement in air & $t^{sep}_{ij}$ & 4.41 s (Set1), 11.79 s (Set2) \\ 
    Wake vortex separation requirement & $t^w_{ij}$ & 0.833 s \\ 
    Occupancy time on TLOF pad & $t^{TOT}_i$ & 2 s \\ 
    OFV ascending/descending time & $t^{R-X}_i, t^{X-R}_i$ & 4.375 s \\ 
    Surface direction ascending and descending time & $t^{X-N}_i, t^{N-X}_i$ & 12.65 s \\ 
    \bottomrule
    \end{tabular}
\end{table}

The time parameters described in the Section~\ref{sec: tc TLOF},~\eqref{eq tAA}-\eqref{eq tda}, are presented in Table~\ref{tab: tlof system}.
\hidetxt{Using these time parameters,} We calculate the maximum time required for \hidetxt{different} all the movement sequence pairs shown in Table~\ref{tab: TLOFpad maxtime}. 
For Set1, the movement pair arrival-arrival or departure-departure gives the maximum achievable throughput ($\mu^{AA}\ or\ \mu^{DD}$) of $( \frac{60}{6.375} )\ 9.41$ VTOLs per minute.
\hidetxt{The maximum throughput of \verti\ $\mu^{\verti}$ \hidetxt{that can be achieved} is $(\frac{60}{19.02})\ 3.15$ VTOLs per minute with only 1 surface direction. 
However, a maximum throughput of 9 VTOLs can be achieved with more than one surface direction.
Therefore, having multiple surface directions significantly increases the throughput of the TLOF pad system.}
Whereas for Set2, the movement pair arrival-arrival or departure-departure gives the maximum achievable throughput ($\mu^{AA}\ or\ \mu^{DD}$) of  $(\frac{60}{11.67})\ 5.14$ VTOLs per minute for single surface direction and $( \frac{60}{6.375} )\ 9.41 $ VTOLs per minute for multiple surface directions.

\begin{table}[h]
\caption{TLOF pad system maximum time calculation for each movement pair}    \label{tab: TLOFpad maxtime}
    \centering
    \begin{tabular}{cccc}
\toprule
\textbf{Set} & \textbf{Surface Directions} & \textbf{Movement Pair} & \textbf{TLOF Times (s)} \\
\hline
 & Single & $T^{AA}_{ij}\ or\ T^{DD}_{ij}$ & 6.375 \\
1 & Single & $T^{AD}_{ij}\ or\ T^{DA}_{ij}$ & 19.02 \\
 & Multiple & $T^{AA}_{ij}\ or\ T^{DD}_{ij}$ & 6.375 \\
 & Multiple & $T^{AD}_{ij}\ or\ T^{DA}_{ij}$ & 6.375 \\
\midrule
 & Single & $T^{AA}_{ij}\ or\ T^{DD}_{ij}$ & 11.67 \\
2 & Single & $T^{AD}_{ij}\ or\ T^{DA}_{ij}$ & 19.02 \\
 & Multiple & $T^{AA}_{ij}\ or\ T^{DD}_{ij}$ & 6.375 \\
 & Multiple & $T^{AD}_{ij}\ or\ T^{DA}_{ij}$ & 6.375 \\
\bottomrule
\end{tabular}
    
\end{table}

\begin{figure}[htbp] 
  \begin{subfigure}[b]{0.52\textwidth}
  \disableFig
    {\includegraphics[width=\textwidth]{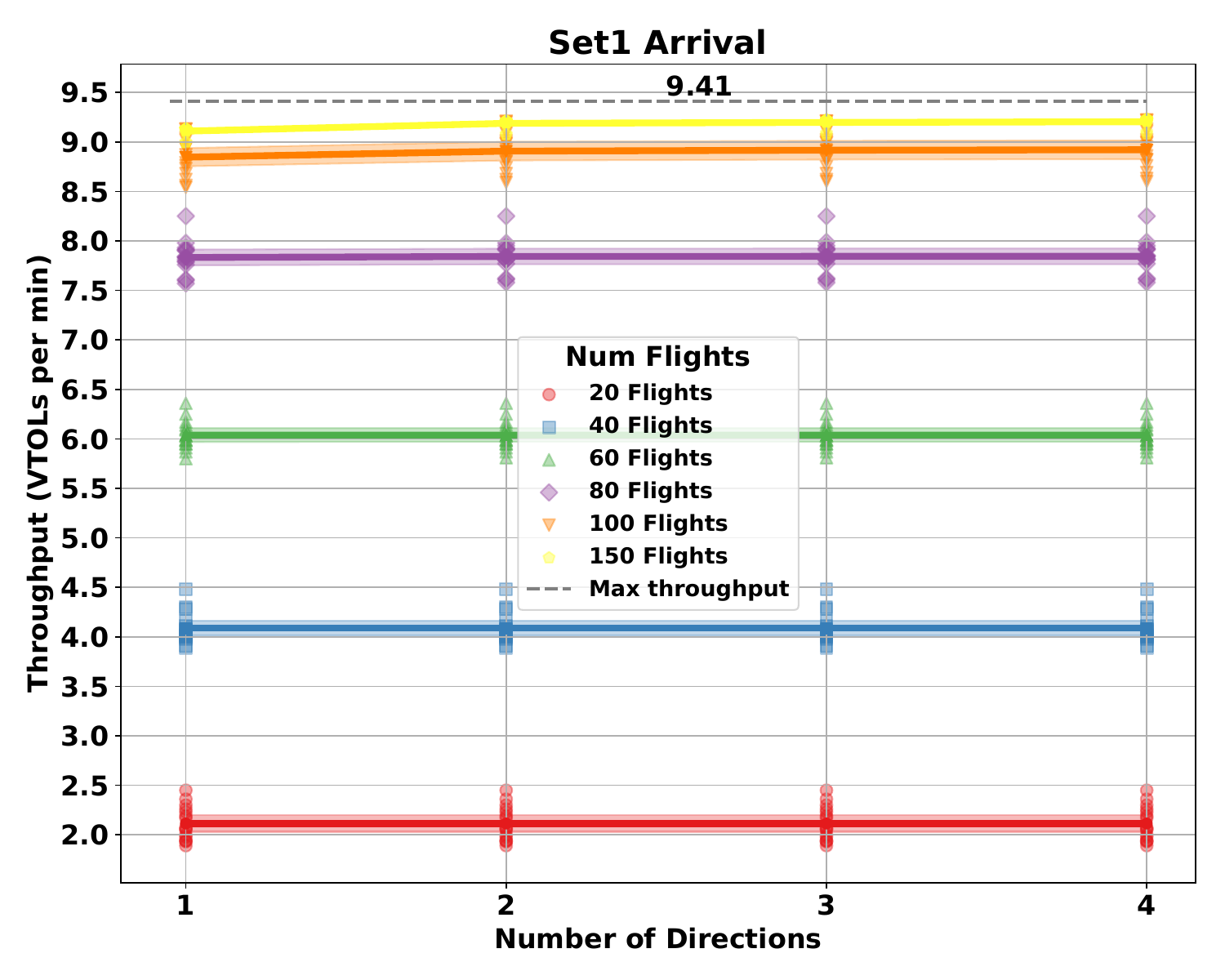}}
    \caption{Arrival}
    \label{fig: throughput_vs_directions_arr}
  \end{subfigure}
  \hfill
  \begin{subfigure}[b]{0.52\textwidth}
    \disableFig
    {\includegraphics[width=\textwidth]{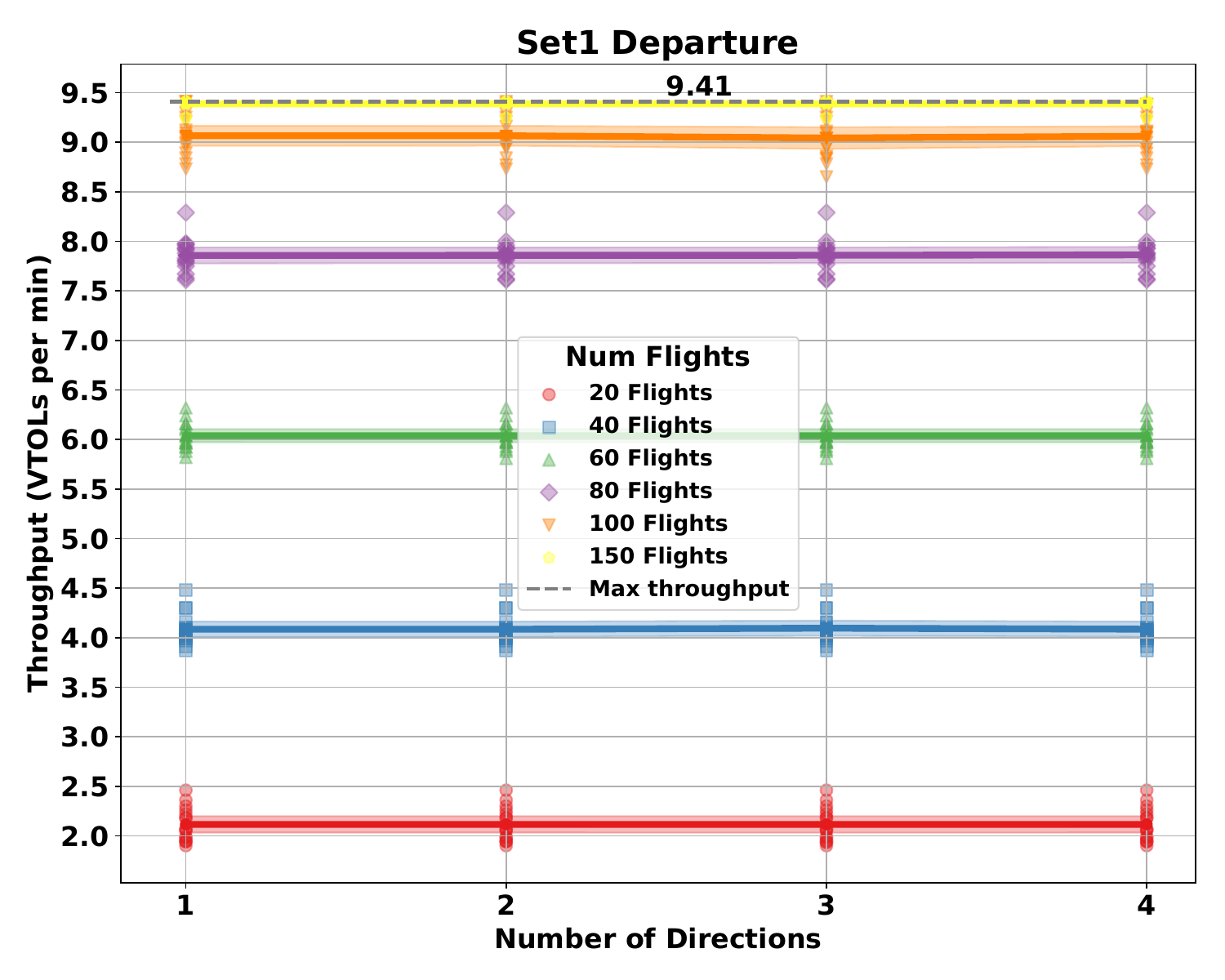}}
    \caption{Departure}
    \label{fig: throughput_vs_directions_dep}
  \end{subfigure}
  \caption{\streva{Set1: Throughput vs Number of directions for a varying number of flights. There is no advantage of multiple surface directions since OFV occupancy time is more than separation time, i.e. $t^{TOT}_i + t^{R-X}_i > t^{sep}_{ij}$} \reva{Set1: Throughput versus the number of directions for varying flight counts. There is no advantage of multiple surface directions since OFV occupancy time exceeds the separation time, i.e. $t^{TOT}_i + t^{R-X}_i > t^{sep}_{ij}$}}
  \label{fig: throughput_vs_directions}
\end{figure}

\begin{figure}[htbp] 
  \begin{subfigure}[b]{0.52\textwidth}
  \disableFig
    {\includegraphics[width=\textwidth]{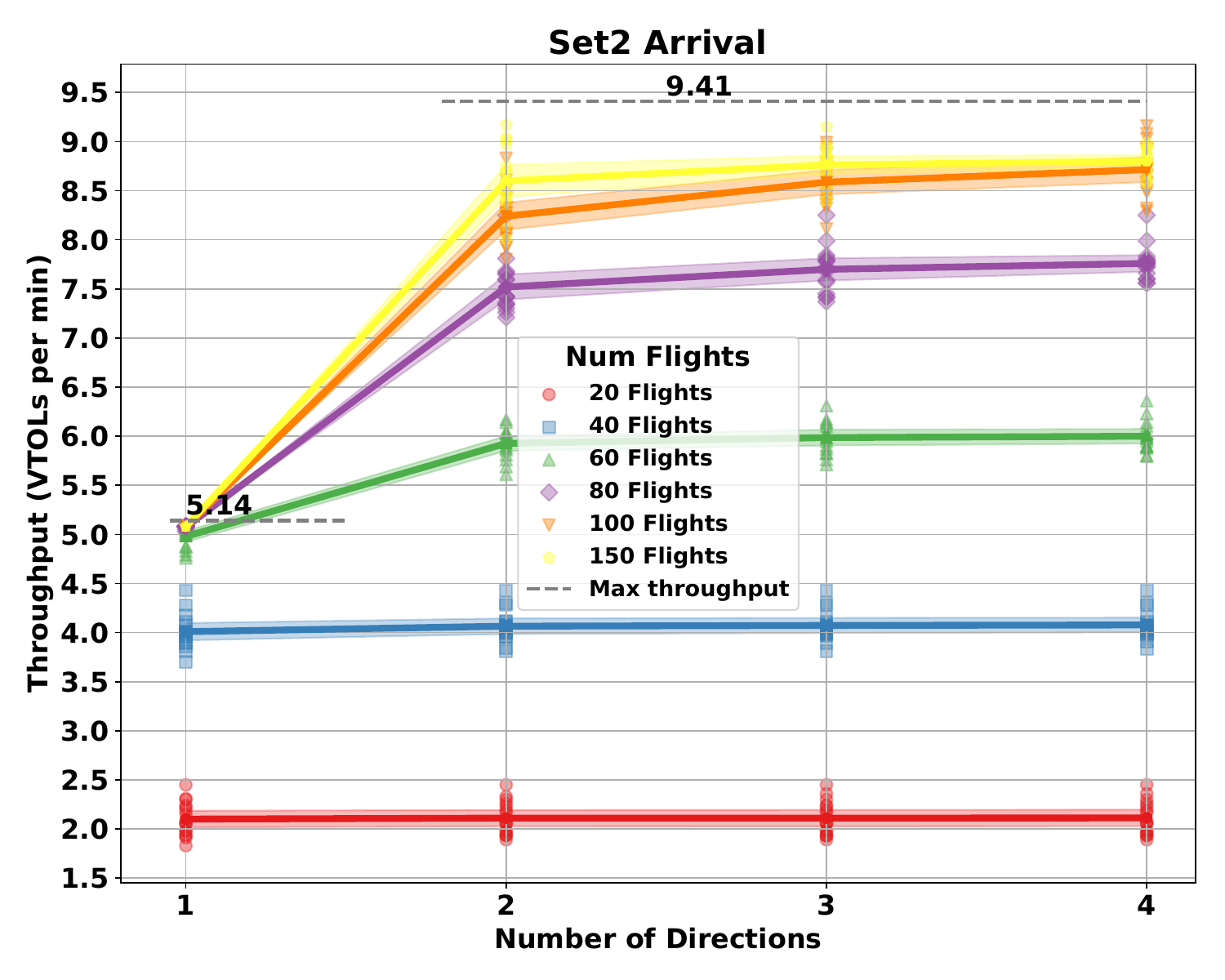}}
    \caption{Arrival}
    \label{fig: throughput_vs_directions_arr2}
  \end{subfigure}
  \hfill
  \begin{subfigure}[b]{0.52\textwidth}
    \disableFig
    {\includegraphics[width=\textwidth]{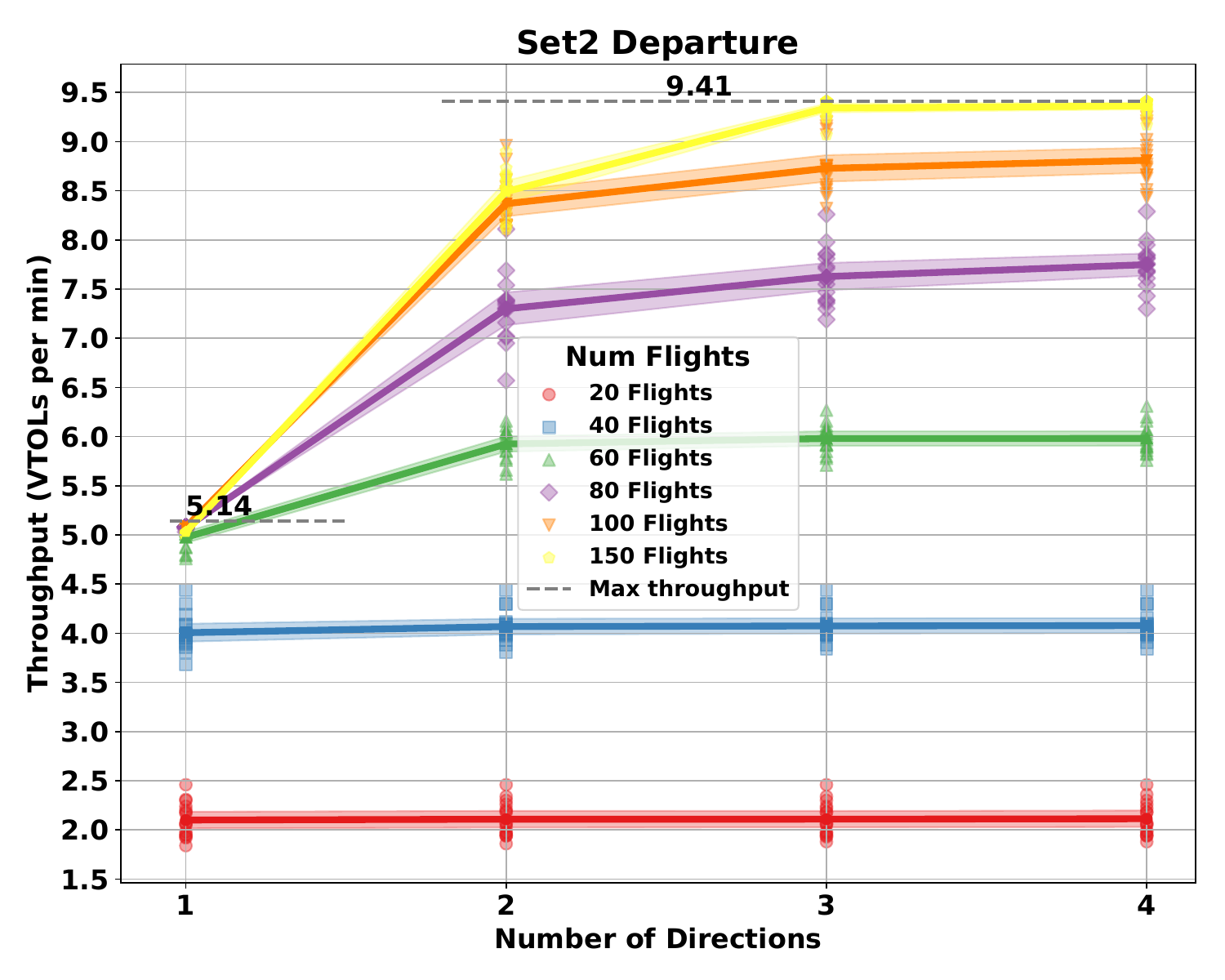}}
    \caption{Departure}
    \label{fig: throughput_vs_directions_dep2}
  \end{subfigure}
  \caption{ \streva{Set2: Throughput vs Number of directions for a varying number of flights. As $t^{TOT}_i + t^{R-X}_i < t^{sep}_{ij}$ the advantage of multiple surface directions is visible as we increase the number of flights} \reva{Set2: Throughput versus the number of directions for varying flight counts. Since the OFV occupancy time is less than the separation time, i.e., $t^{TOT}_i + t^{R-X}_i < t^{sep}_{ij}$, the benefit of multiple surface directions becomes evident as the flight count increases.}}
  \label{fig: throughput_vs_directions2}
\end{figure}

If two consecutive VTOLs have different routes, then \hidetxt{multiple surface directions cause} the separation requirement for the same surface direction, $t^{sep}_{ij}$, is set to 0.
In Set1, since the combined TLOF pad occupancy time and OFV travel time ($t^{TOT}_i+t^{R-X}$) is greater than $t^{sep}_{ij}$, there is no advantage of multiple surface directions.
Conversely, in Set2,  $t^{TOT}_i+t^{R-X} < t^{sep}_{ij}$ and thus having multiple surface directions gives a higher throughput.
The same can be observed in Figures~\ref{fig: throughput_vs_directions}~-~\ref{fig: throughput_vs_directions2}, which show the throughput achieved by solving the MILP formulation~\eqref{obj eqn}. 
For the \textit{TLOF pad system}, about 400 MILP computations were run with various sets of flight numbers and surface directions. 
Each set of computations involved different random arrival and departure timings. 
To observe the maximum throughput, the input number of flights needs to be significantly high. 
The plots of 100 and 150 flights thus validate both our MILP formulation and the throughput calculations, including the advantage of multiple surface directions for Set2, as mentioned earlier.

\newcommand{\scalesz}{0.5}

\subsubsection{Gate system}
Recall the topology shown in Figure \ref{fig:topology}, which has 4 gates with 3 parking slots each, resulting in a static capacity ($\mathbb{C}_G$) of 12 VTOLs.
 The maximum dynamic capacity (throughput, $\mu^G$) depends on the turnaround time ($T^{TAT}$):
\begin{enumerate}
    \item For Set1, $\mu^G = \frac{60 \times 12}{90} = 8$ VTOLs per minute.
    \item For Set2, $\mu^G = \frac{60 \times 12}{120} = 6$ VTOLs per minute.
\end{enumerate}

\begin{figure}[h!] 
  \begin{subfigure}[b]{0.5\textwidth}
  \disableFig
    {
    \includegraphics[width=1\textwidth]{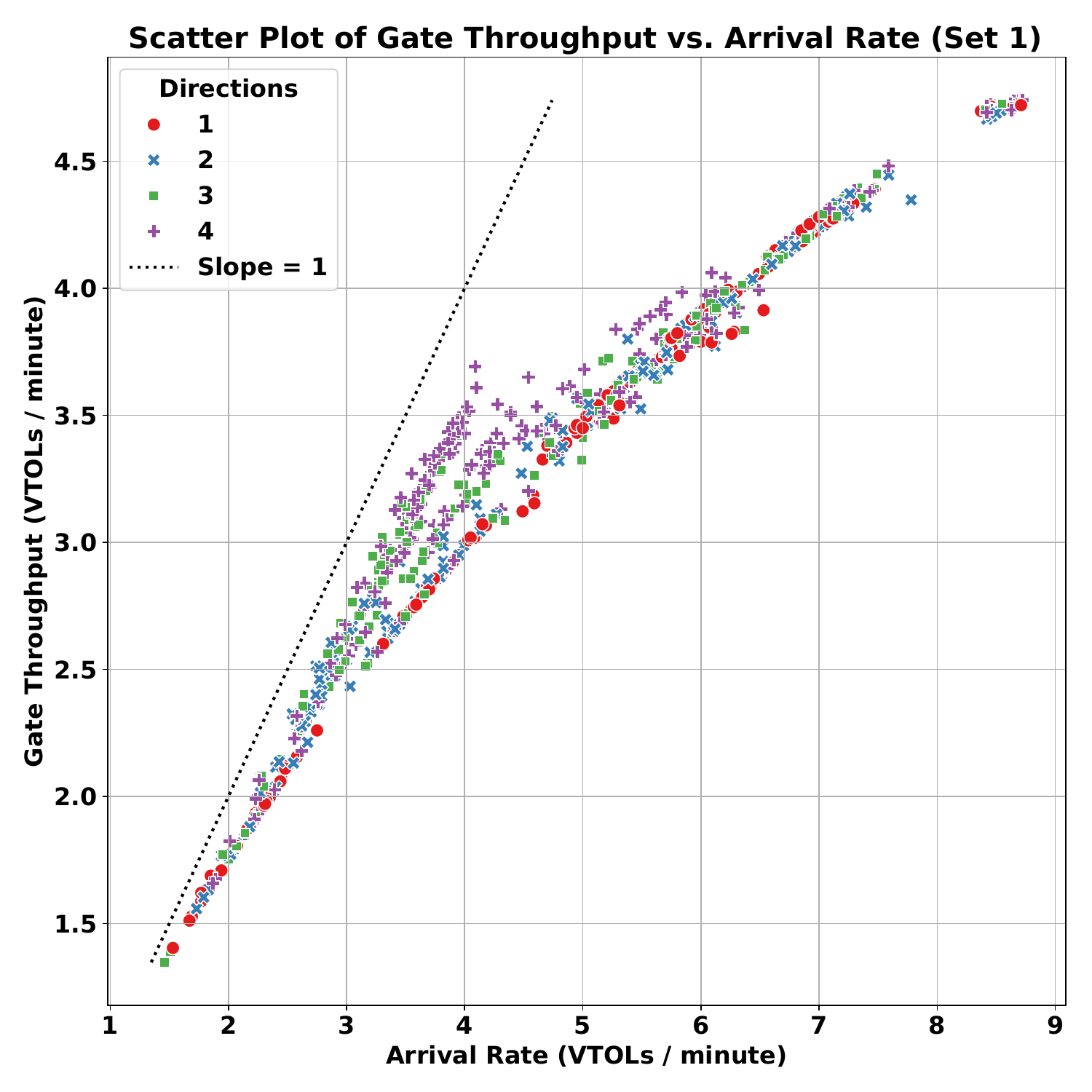}
    }
    \caption{Set1, $t^{TOT}_i+t^{R-X}_i > t^{sep}_{ij}$ and $T^{TAT}=90s$}
    \label{fig: gate rate set1}
  \end{subfigure}
  \hfill
  \begin{subfigure}[b]{0.5\textwidth}
    \disableFig
    {
    \includegraphics[width=1\textwidth]{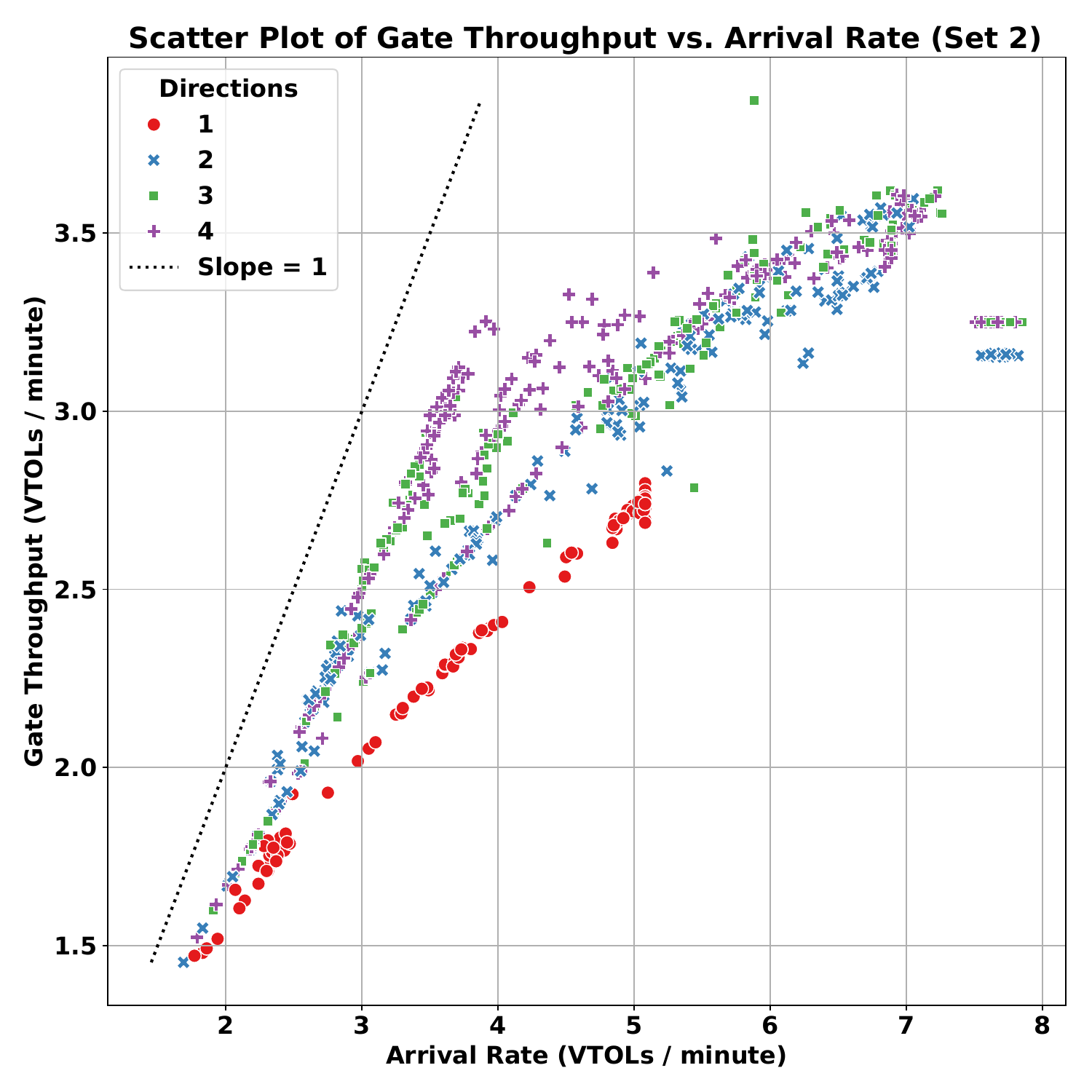}
    }
    \caption{Set2, $t^{TOT}_i + t^{R-X}_i < t^{sep}_{ij}$ and $T^{TAT}=120s$}
    \label{fig: gate rate set2}
  \end{subfigure}
  \caption{\reva{Gate system throughput versus arrival rate. Figure~\ref{fig: gate rate set1} (Set1) depicts conditions where OFV occupancy time is more than separation time and turnaround time ($T^{TAT}$) is 90 seconds, while in Figure~\ref{fig: gate rate set2} (Set2), OFV occupancy time is less than separation time and $T^{TAT}$ is 120 seconds. The Slope=1 line confirms that gate throughput remains consistently below arrival throughput. Notably, the gate throughput decreases as $T^{TAT}$ increases, as evidenced by the higher throughput in Set1 compared to Set2. However, the arrival rate is bounded by time parameters, as further illustrated in Figure~\ref{fig: gate rate set2} for single directions.}}
  \label{fig: gate rate}
\end{figure}

\hidetxt{However, these are steady-state values assuming immediate slot refilling. 
In the given topology and computation, this assumption may not hold; hence, we show the maximum rate of a single gate slot($1/T^{TAT}$), i.e. 0.667 VTOLS per minute for Set1 and 0.5 VTOLs per minute for Set2.
Figures~\ref{fig: gate rate} show the variations of gate slot throughput, with arrival rate for different numbers of directions obtained by solving approximately 1000 MILP for each of the 2 sets using turn around flights ($A^{TAT}$). 
The results indicate:
\begin{itemize}
    \item Maximum slot throughput is achievable in some cases, irrespective of the number of directions.
    \item Slot throughput is consistently lower than the arrival rate, as expected.
\end{itemize}
These findings validate that the maximum gate rate depends solely on $T^{TAT}$, while the number of directions influences only the arrival rate. This is further demonstrated in Figure~\ref{fig: gate rate set2}.}

\reva
{
Figures~\ref{fig: gate rate} illustrate variations in gate throughput with arrival rates, based on approximately 1,000 MILP solutions for each set using turnaround flights ($A^{TAT}$). Key observations include:
\begin{itemize}
    \item \textit{The gate throughput increases with lower turnaround time ($T^{TAT}$)}\hidetxt{The gate throughput is primarily affected by turn-around time.}, as in Set1.
    \item As discussed in previous subsection~\ref{Sec: TLOF pad sys results}, \textit{Arrival rate is bounded by time parameters} as can be seen from Figure~\ref{fig: gate rate set2} for single surface direction where OFV occupancy time is less than separation time.
    \item As expected, \textit{the gate throughput is consistently lower than the arrival rate}, with all data points falling below the line with slope = 1.
\end{itemize}
}

\subsubsection{Taxiway system}
Using the~\eqref{eq taxi flow} from Section~\ref{sec: tc taxi}, the maximum flow rate of the taxiway link is calculated to be $(\frac{60}{(10/6)}) 36$ VTOLs per minute, which \hidetxt{is the maximum flow rate of}represents the entire taxiway network.
This \hidetxt{taxiway flow rate} is significantly higher than the maximum throughput of the TLOF pad system and Gate system, indicating that taxiway flow rate is not a limiting factor for the \verti\ capacity.

\subsubsection{\rev{Vertiminal Throughput}}
Finally, We obtain the \verti's maximum throughput using~\eqref{eq tc verti}. For Set1, it is 8 VTOLs per minute, while for Set2, it is 5.14 VTOLs per minute for single surface direction and 6 VTOLs per minute for multiple surface directions.

\hidetxt{No significant difference in throughput was observed between different number of surface directions due to the consideration of only `Small' VTOLs In the Figures, there is no visible change observed between multiple surface directions as we have only considered ``Small" VTOLs in which occupancy time and time to cover OFV ($T^{TOT}+t^{R-X}$) is more than the separation requirement on a surface direction ($t^{sep}_{ij}$).}
\hidetxt{Multiple surface directions may alleviate congestion for departing. However, they can increase the congestion at the OFV for the arrivals.}

Our throughput results also align with other works mentioned in \cite{vascik2019development}. 
The authors consider only consolidated arrival and departure times without any consideration of separation requirements. 
In contrast, from Section~\ref{sec: tc TLOF}, for a single VTOL, arrival time = $t^{N-X}_i + t^{X-R}_i + t^{TOT}_i$ and departure time = $ t^{TOT}_i + t^{R-X}_i + t^{X-N}_i$, shows in detail the occupancy time of TLOF pad, OFV, and travel time on a surface direction.
Furthermore, from~\eqref{eq tAA} - \eqref{eq tda}, the TLOF pad throughput is affected by the maximum of individual time parameters rather than the consolidated arrival or departure time. 
By increasing any time parameter, arrival and/or departure time may increase, which decreases the throughput, thus validating the results shown in  \cite{vascik2019development}. 
Consequently, the estimation of $\mathbb{C}_G$ required for a single TLOF pad \hidetxt{also gets affected,} is shown in~\eqref{eq gate cap}.

\begin{equation}
    \mu^{G} = \mu^{TLOF} \\
    => \mathbb{C}_G = \frac{TAT}{\min(T^{AA}_{ij}, T^{DD}_{ij})}
    \label{eq gate cap}
\end{equation}

Unlike \cite{vascik2019development}, our equations allow a detailed analysis of throughput of the TLOF pad system ($\mu^{TLOF}$), taxiway system ($\mu^{taxiway}$) and the gate system  ($\mu^{G}$), and thus enables deeper understanding \hidetxt{of the impact of a bottleneck} and the significance of the TLOF pad to gate ratio. 

\section{Case Study: Application to Gimpo Vertiminal} \label{sec: Case Study} 

\begin{figure}
    \centering
    \includegraphics[width=0.8\linewidth]{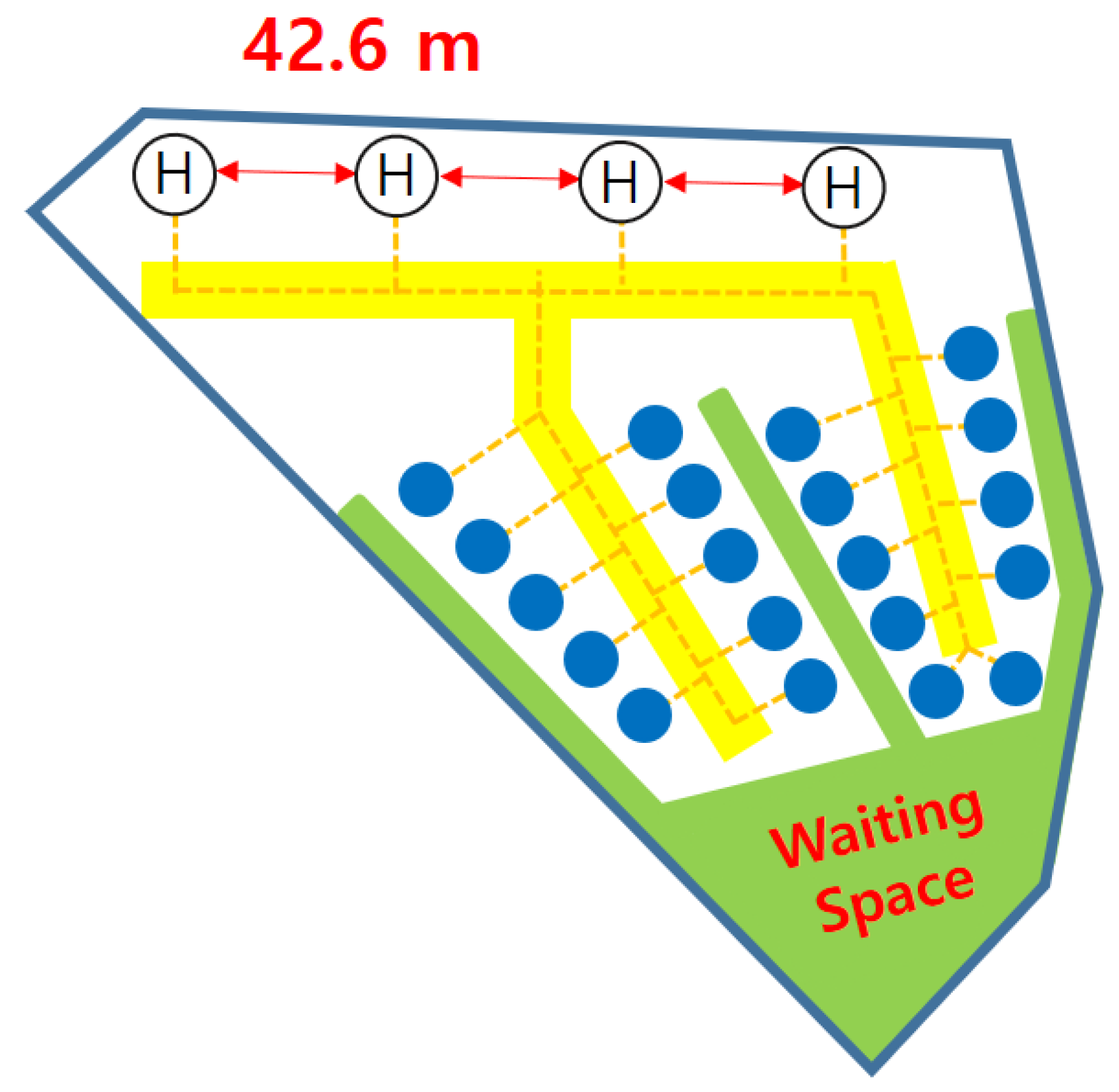}
    \caption{Gimpo Vertiminal topology given by~\cite{ahn2022design}}
    \label{fig: gimpo topo}
\end{figure}

The \textbf{aim} of this case study is to validate the effectiveness of our proposed optimisation framework and throughput analysis by applying it to a realistic vertiminal design. 
We also seek to demonstrate how different operational configurations impact the overall throughput of a vertiminal and to identify the configuration that maximises throughput using our MILP formulation.
By analysing the vertiminal design proposed for Gimpo Airport~\cite{ahn2022design}, Korea, we aim to provide quantitative insights into UAM infrastructure planning and showcase the practical utility of our framework.
Gimpo Airport in Korea is a major hub for domestic flights and has been proposed as a candidate for integrating UAM services, presenting a realistic scenario for applying our framework.
Specifically, we selected the pier-type topology featuring four TLOF pads and twenty gates, adhering to EASA guidelines, as shown in Figure~\ref{fig: gimpo topo}. 
This design aims to maximise TLOF pad and gate utilisation.

\subsection{Setup}
\begin{figure}
    \centering
    \includegraphics[width=0.8\linewidth]{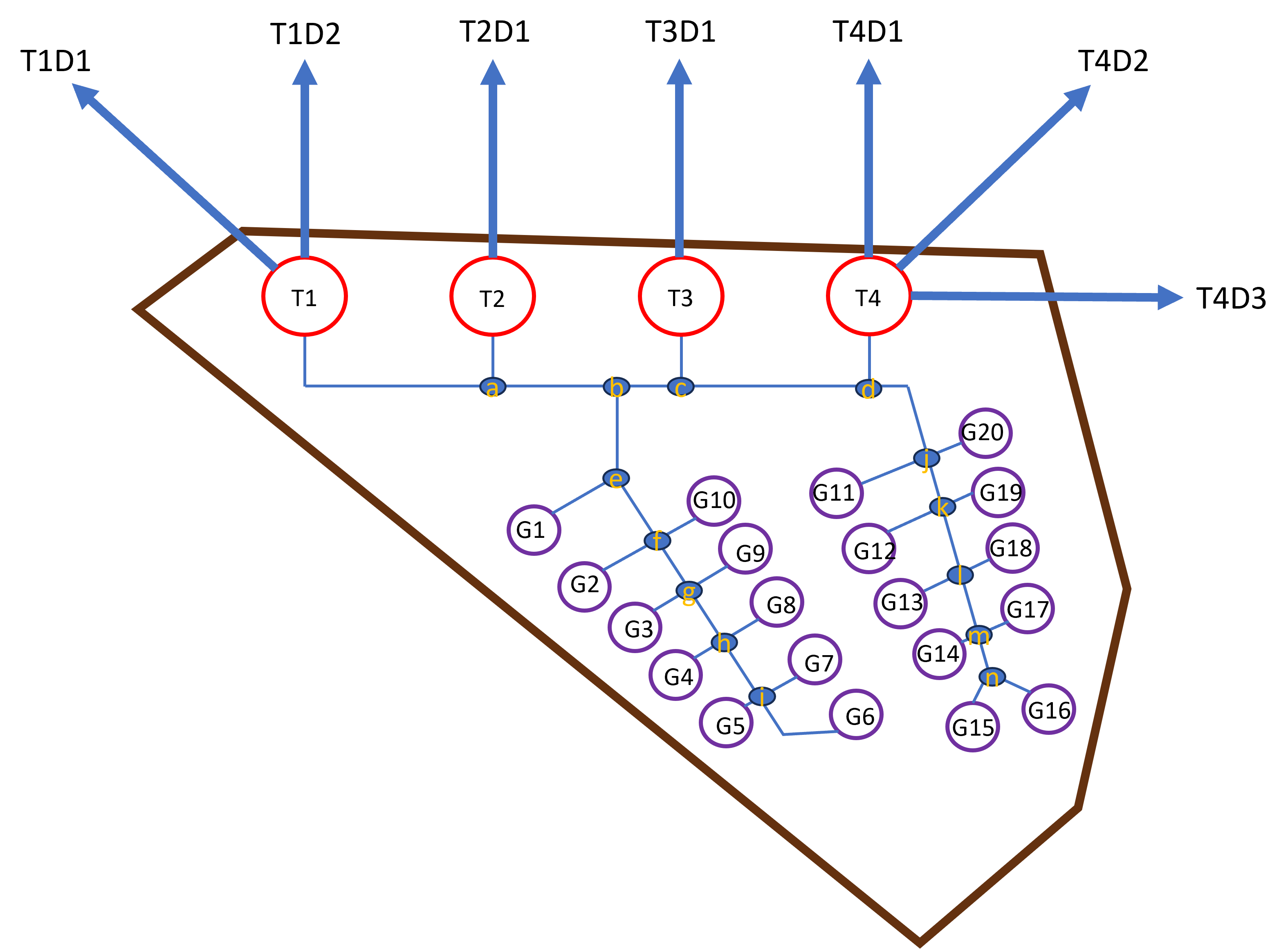}
    \caption{Gimpo Vertiminal topology with nodes(4 TLOF pads and 20 gates) and directions}
    \label{fig: gimpo topo dir}
\end{figure}

We evaluated four different operational configurations of the vertiminal on the Gimpo topology shown in Figure~\ref{fig: gimpo topo dir}:
\begin{enumerate}
    \item \textbf{Configuration 1}: All four TLOF pads are used for both arrivals and departures for all gates, representing a fully flexible operation with potential for high congestion due to shared usage.
    \item \textbf{Configuration 2}: TLOF pads T1–T2 serve arrivals and departures for gates G1–G10, while T3–T4 serve gates G11–G20, introducing segmentation to reduce path conflicts between the two parts of the terminal.
    \item \textbf{Configuration 3}: TLOF pads T1 and T4 are designated for departures, and T2–T3 for arrivals, serving all gates. This segregation aims to minimise interactions between arrivals and departures across all TLOF pads.
    \item \textbf{Configuration 4}: TLOF pads T1 and T4 are designated for arrivals, and T2–T3 for departures, serving all gates, providing an alternative segregation strategy to reduce congestion.
\end{enumerate}

To study throughput bounds, for each configuration, we applied the vertiminal parameters shown in Table~\ref{tab: param} and two sets of operational parameters shown in Table~\ref{tab: tlof system} from earlier sections. 
Since we analysed the TLOF pads system, taxiways system, and gates system in isolation, for a particular operational parameter set, the theoretical throughput bounds are constant across all configurations within that parameter set.
However, these bounds vary between the two operational parameter sets, as shown in Table~\ref{tab: gimpo throughput}.

\begin{table}[h]
\centering
\caption{Maximum throughput bounds for Set 1 and Set 2 in terms of vehicles/minute.}\label{tab: gimpo throughput}
\begin{tabular}{lcc}
\toprule
\textbf{Parameter} & \textbf{Set 1} & \textbf{Set 2} \\
\midrule
$\mu^{TLOF}$ & 37  & 20  \\
$\mu^{taxi}$ & 36  & 36  \\
$\mu^G$      & 13 & 10  \\
\midrule
$\mu^{vertiminal}$ & 13  & 10  \\
\bottomrule
\end{tabular}
\end{table}

We executed approximately 1,000 MILP computations for each configuration and parameter set, varying seed values and the number of scheduled flights. 

\subsection{Insights}

The observed maximum throughput, summarised in Table~\ref{tab: gimpo opt throughput}, approached but did not exceed the theoretical bounds as expected.
This discrepancy stems from analysing system elements independently and assuming idealised conditions, such as instantaneous vehicle availability at a node and congestion-free paths.
Configurations 3 and 4, which segregate arrivals and departures, consistently achieved higher throughput compared to Configurations 1 and 2. 
The segregation of TLOF pads for arrivals and departures reduces path conflicts \hidetxt{and minimises interactions between vehicles, }enhancing operational efficiency.
Configuration 2 has the lowest throughput as segmentation inadvertently increases the path conflicts within each segment. 
The use of a single path from multiple gates to the TLOF pads creates bottlenecks, limiting the ability to manage vehicle flow efficiently.
These results highlight the importance of operational strategies that reduce path conflicts and congestion of UAM vehicles.
By strategically allocating TLOF pads for arrival and departure and optimising vehicle movement paths, vertiport operators can significantly enhance throughput.

\begin{table}[h]
\centering
\caption{Maximum throughput of configurations achieved in vehicles/minute.}\label{tab: gimpo opt throughput}
\begin{tabular}{lcc}
\toprule
\textbf{Configuration} & \textbf{Set 1} & \textbf{Set 2} \\
\midrule
Configuration 1 & 9.13 & 8.79 \\
Configuration 2 & 8.88 & 8.21 \\
Configuration 3 & 9.25 & 9.09 \\
Configuration 4 & 9.54 & 9.12 \\
\bottomrule
\end{tabular}
\end{table}

\subsection{Summary}
This case study validates the robustness of our optimisation framework and confirms its utility for real-world UAM infrastructure planning. 
By applying our MILP formulation to different operational configurations, we demonstrated that the choice of configuration has a significant impact on throughput. 
Specifically, configurations that segregate arrivals and departures (Configurations 3 and 4) outperform those that do not.

Our findings make it explicit that using our MILP formulation, we can choose the best configuration for maximum throughput. 
By adopting configurations that strategically allocate TLOF pads for arrivals and departures, vertiport operators can enhance capacity and reduce delays, facilitating efficient UAM traffic. 
This optimisation capability is crucial for meeting the demands of future urban air mobility networks and ensuring the smooth integration of UAM services into existing transportation infrastructure.

\section{Conclusion}
\label{conclusion}

\nonanon{This paper builds upon our prior work~\cite{saxena2023integrated} by extending the optimization framework}
\anon{This paper extends the optimization framework presented in~\cite{saxena2023integrated}}
to include arrival flights and turnaround at gates, enabling the generation of optimal schedules for a wider range of vertiport terminal (vertiminal) operations such as arrival-departure and departure-arrival. 
Further improvements to the Mixed Integer Linear Program (MILP) formulation has drastically reduced both the formulation and solver time.
Our formulation can achieve up to 50\% delay reduction compared to First-Come, First-Serve (FCFS) scheduling. 
This performance gain stems from efficiently sequencing VTOL classes and utilizing multiple surface directions. 

Our work delves deeper into \verti's throughput by holistically analyzing its three core elements: the TLOF pad system, taxiway network, and gate system. 
In the case of TLOF pad system, our computation results demonstrate that when flight separation time exceeds OFV occupancy time, the multiple surface directions greatly enhance throughput. 
On the other hand, gate system throughput is constrained by longer turnaround times, as validated by the MILP formulation. 
Taxiway network is usually not a bottleneck to the vertiminal throughput.
Thus, by deriving precise throughput equations, we provide practitioners with analytical tools to optimize vertiminal design and operations without relying on computationally expensive simulations.
Using our framework on one of the most efficient vertiminal designs at Gimpo~\cite{ahn2022design}, we demonstrated that the MILP formulation can effectively identify optimal operational configurations. 

For future work, we aim to further reduce MILP computation time by incorporating scalable techniques such as rolling horizon methods.
We also plan to develop stochastic optimisation and data-driven reinforcement learning models to better handle operational uncertainties and enhance Quality of Service (QoS) metrics, including equitable delay distribution and position inversion constraints.
Future efforts will also involve creating and analysing a capacity envelope for vertiports to better understand their operational limits under various configurations.

\bibliographystyle{unsrt}
\bibliography{bibliography}

\begin{thebibliography}{10}

\bibitem{9447255}
Adam~P. Cohen, Susan~A. Shaheen, and Emily~M. Farrar.
\newblock Urban air mobility: History, ecosystem, market potential, and challenges.
\newblock {\em IEEE Transactions on Intelligent Transportation Systems}, 22(9):6074--6087, 2021.

\bibitem{nasaConOps}
NASA.
\newblock Uam vision concept of operations (conops) uam maturity level (uml) 4.

\bibitem{McKinsey}
McKinsey.
\newblock Future air mobility: Major developments in 2022 and significant milestones ahead, 2023.

\bibitem{volo1}
Hayley Everett.
\newblock Italy's first vertiport deployed at fiumicino airport, 2022.

\bibitem{volo2}
FutureFlight.
\newblock Volocopter concludes european urban air mobility airspace integration flight trials, 2022.

\bibitem{skyroad}
FutureFlight.
\newblock Advanced air mobility flight test site to open at german airport, 2023.

\bibitem{9441631}
Mihir Rimjha and Antonio Trani.
\newblock Urban air mobility: Factors affecting vertiport capacity.
\newblock In {\em 2021 Integrated Communications Navigation and Surveillance Conference (ICNS)}, pages 1--14, 2021.

\bibitem{EASADoc}
EASA.
\newblock {\em Prototype Technical Specifications for the Design of VFR Vertiports for Operation with Manned VTOL-Capable Aircraft Certified in the Enhanced Category}.

\bibitem{FAAdoc}
FAA.
\newblock {\em ENGINEERING BRIEF \#105 Vertiport Design}.

\bibitem{UAE}
UAE General Civil~Aviation Authority, 2022.

\bibitem{Watkins9473838}
Lanier Watkins, Nick Sarfaraz, Sebastian Zanlongo, Joshua Silbermann, Tyler Young, and Randall Sleight.
\newblock An investigative study into an autonomous uas traffic management system for congested airspace safety.
\newblock In {\em 2021 IEEE International Conference on Communications Workshops (ICC Workshops)}, pages 1--6, 2021.

\bibitem{mueller2017enabling}
Eric~R Mueller, Parmial~H Kopardekar, and Kenneth~H Goodrich.
\newblock Enabling airspace integration for high-density on-demand mobility operations.
\newblock In {\em 17th AIAA Aviation Technology, Integration, and Operations Conference}, page 3086, 2017.

\bibitem{jiang2015taxiing}
Yu~Jiang, Xinxing Xu, Honghai Zhang, and Yuxiao Luo.
\newblock Taxiing route scheduling between taxiway and runway in hub airport.
\newblock {\em Mathematical Problems in Engineering}, 2015, 2015.

\bibitem{lee2012comparison}
Hanbong Lee and Hamsa Balakrishnan.
\newblock A comparison of two optimization approaches for airport taxiway and runway scheduling.
\newblock In {\em 2012 IEEE/AIAA 31st Digital Avionics Systems Conference (DASC)}, pages 4E1--1. IEEE, 2012.

\bibitem{deng2020novel}
Wu~Deng, Junjie Xu, Huimin Zhao, and Yingjie Song.
\newblock A novel gate resource allocation method using improved pso-based qea.
\newblock {\em IEEE Transactions on Intelligent Transportation Systems}, 2020.

\bibitem{vazquez2021vertiport}
Hack V\'azquez et~al.
\newblock Vertiport sizing and layout planning through integer programming in the context of urban air mobility, 2021.

\bibitem{zelinski2020operational}
Shannon Zelinski.
\newblock Operational analysis of vertiport surface topology.
\newblock In {\em 2020 AIAA/IEEE 39th Digital Avionics Systems Conference (DASC)}, pages 1--10. IEEE, 2020.

\bibitem{shao2021terminal}
Quan Shao, Mengxue Shao, and Yang Lu.
\newblock Terminal area control rules and evtol adaptive scheduling model for multi-vertiport system in urban air mobility.
\newblock {\em Transportation Research Part C: Emerging Technologies}, 132:103385, 2021.

\bibitem{song2021development}
Kyowon Song and Hwasoo Yeo.
\newblock Development of optimal scheduling strategy and approach control model of multicopter vtol aircraft for urban air mobility (uam) operation.
\newblock {\em Transportation research Part C: emerging technologies}, 128:103181, 2021.

\bibitem{furini2015improved}
Fabio Furini, Martin~Philip Kidd, Carlo~Alfredo Persiani, and Paolo Toth.
\newblock Improved rolling horizon approaches to the aircraft sequencing problem.
\newblock {\em Journal of Scheduling}, 18:435--447, 2015.

\bibitem{saxena2023integrated}
Ravi~Raj Saxena, Tejas Joshi, DK~Yashashav, TV~Prabhakar, and Joy Kuri.
\newblock Integrated taxiing and tlof pad scheduling using different surface directions with fairness analysis.
\newblock In {\em 2023 IEEE 26th International Conference on Intelligent Transportation Systems (ITSC)}, pages 1747--1752. IEEE, 2023.

\bibitem{ahn2022design}
Byeongseon Ahn and Ho-Yon Hwang.
\newblock Design criteria and accommodating capacity analysis of vertiports for urban air mobility and its application at gimpo airport in korea.
\newblock {\em Applied Sciences}, 12(12):6077, 2022.

\bibitem{gotteland2003genetic}
J-B Gotteland and Nicolas Durand.
\newblock Genetic algorithms applied to airport ground traffic optimization.
\newblock In {\em The 2003 Congress on Evolutionary Computation, 2003. CEC'03.}, volume~1, pages 544--551. IEEE, 2003.

\bibitem{liu2010airport}
Changyou Liu and Kaifeng Guo.
\newblock Airport taxi scheduling optimization based on genetic algorithm.
\newblock In {\em 2010 International Conference on Computational Intelligence and Security}, pages 205--208. IEEE, 2010.

\bibitem{clare2011optimization}
Gillian Clare and Arthur~G Richards.
\newblock Optimization of taxiway routing and runway scheduling.
\newblock {\em IEEE Transactions on Intelligent Transportation Systems}, 12(4):1000--1013, 2011.

\bibitem{simaiakis2013analysis}
Ioannis Simaiakis.
\newblock {\em Analysis, modelling and control of the airport departure process}.
\newblock PhD thesis, Massachusetts Institute of Technology, 2013.

\bibitem{cheng2014airport}
Peng Cheng, Xiang Zou, and Wenda Liu.
\newblock Airport surface trajectory optimization considering runway exit selection.
\newblock In {\em 17th International IEEE Conference on Intelligent Transportation Systems (ITSC)}, pages 2656--2662. IEEE, 2014.

\bibitem{herrema2019machine}
Floris Herrema, Ricky Curran, Sander Hartjes, Mohamed Ellejmi, Steven Bancroft, and Michael Schultz.
\newblock A machine learning model to predict runway exit at vienna airport.
\newblock {\em Transportation Research Part E: Logistics and Transportation Review}, 131:329--342, 2019.

\bibitem{morgan2019validation}
Catherine~Chalon Morgan, Mohamed Ellejmi, Floris Herrema, and Ricky Curran.
\newblock Validation of the runway utilisation concept.
\newblock {\em 9th SESAR Innovation Days}, 2019.

\bibitem{behrends2016aircraft}
John~A Behrends and John~M Usher.
\newblock Aircraft gate assignment: using a deterministic approach for integrating freight movement and aircraft taxiing.
\newblock {\em Computers \& Industrial Engineering}, 102:44--57, 2016.

\bibitem{deng2017study}
Wu~Deng, Meng Sun, Huimin Zhao, Bo~Li, and Chunxiao Wang.
\newblock Study on an airport gate assignment method based on improved aco algorithm.
\newblock {\em Kybernetes}, 47(1):20--43, 2017.

\bibitem{kleinbekman2018evtol}
Imke~C Kleinbekman, Mihaela~A Mitici, and Peng Wei.
\newblock evtol arrival sequencing and scheduling for on-demand urban air mobility.
\newblock In {\em 2018 IEEE/AIAA 37th Digital Avionics Systems Conference (DASC)}, pages 1--7. IEEE, 2018.

\bibitem{vascik2019development}
Parker~D Vascik and R~John Hansman.
\newblock Development of vertiport capacity envelopes and analysis of their sensitivity to topological and operational factors.
\newblock In {\em AIAA Scitech 2019 Forum}, page 0526, 2019.

\bibitem{preis2022vertiport}
Lukas Preis and Mirko Hornung.
\newblock A vertiport design heuristic to ensure efficient ground operations for urban air mobility.
\newblock {\em Applied Sciences}, 12(14):7260, 2022.

\bibitem{bertsimas2000traffic}
Dimitris Bertsimas and Sarah~Stock Patterson.
\newblock The traffic flow management rerouting problem in air traffic control: A dynamic network flow approach.
\newblock {\em Transportation Science}, 34(3):239--255, 2000.

\bibitem{tsao2009integrated}
H-S~Jacob Tsao, Wenbin Wei, Agus Pratama, and JR~Tsao.
\newblock Integrated taxiing and take-off scheduling for optimization of airport surface operations.
\newblock In {\em Proc. 2nd Annual Conference of Indian Subcontinent Decision Science Institute (ISDSI 2009)}, pages 3--5, 2009.

\bibitem{RRJTOSN}
Ravi~Raj Saxena, Joydeep Pal, Srinivasan Iyengar, Bhawana Chhaglani, Anurag Ghosh, Venkata~N. Padmanabhan, and Prabhakar~T. Venkata.
\newblock Holistic energy awareness and robustness for intelligent drones.
\newblock {\em ACM Trans. Sen. Netw.}, jan 2024.
\newblock Just Accepted.

\bibitem{de2020airport}
Richard De~Neufville.
\newblock Airport systems planning, design, and management.
\newblock In {\em Air Transport Management}, pages 79--96. Routledge, 2020.

\bibitem{mfmc}
Tom Leighton and Satish Rao.
\newblock Multicommodity max-flow min-cut theorems and their use in designing approximation algorithms.
\newblock {\em J. ACM}, 46(6):787–832, nov 1999.

\bibitem{github}
github.
\newblock https://github.com/rrsaxena92/Vertiport\_schedule/.

\end{thebibliography}

\end{document}